\documentclass[floats,twocolumn,aps,prb,longbibliography,superscriptaddress]{revtex4-2}

\usepackage{graphicx}
\usepackage{amsfonts}
\usepackage{amssymb}
\usepackage{dsfont}
\usepackage{amsbsy}
\usepackage[colorlinks=true, urlcolor=blue, linkcolor=blue, citecolor=blue, pdftex]{hyperref}
\usepackage{hyperref}
\usepackage{xcolor}
\usepackage{subeqnarray}
\usepackage{tabularx}
\usepackage{tabularray}
\usepackage{fancyhdr}
\usepackage[version=4]{mhchem}
\newcommand{\nab}{\ce{KCu7TeO4(SO4)5Cl}}
\newcommand{\Nanab}{\ce{NaCu7TeO4(SO4)5Cl}}
\newcommand{\Rbnab}{\ce{RbCu7TeO4(SO4)5Cl}}
\newcommand{\Csnab}{\ce{CsCu7TeO4(SO4)5Cl}}
\newcommand{\skag}{\ce{Na6Cu7BiO4(PO4)4Cl3}}
\newcommand{\Alatla}{\ce{KCu6AlBiO4(SO4)5Cl}}
\newcommand{\atla}{\ce{KCu6FeBiO4(SO4)5Cl}}

\usepackage{placeins}
\newcommand{\beginsupplement}{%
    \setcounter{table}{0}
    \renewcommand{\thetable}{\arabic{table}}%
    \renewcommand{\tablename}{Supplementary Table}
    \setcounter{figure}{0}
    \renewcommand{\figurename}{Supplementary Figure}
    \setcounter{equation}{0}
    \setcounter{page}{1}
    \setcounter{section}{0}
    \renewcommand{\thesection}{Supplementary Note \arabic{section}}%
}

\allowdisplaybreaks[1]

\begin{document}

\title{Field-induced spin liquid in the decorated square-kagome antiferromagnet nabokoite {\nab}}

\author{Mat\'ias G. Gonzalez}
\email{matias.gonzalez.phys@gmail.com}
\affiliation{Helmholtz-Zentrum Berlin f\"ur Materialien und Energie, Hahn-Meitner-Platz 1, 14109 Berlin, Germany}
\affiliation{Dahlem Center for Complex Quantum Systems and Fachbereich Physik, Freie Universität Berlin, 14195 Berlin, Germany\looseness=-1}

\author{Yasir Iqbal}
\email{yiqbal@physics.iitm.ac.in}
\affiliation{Department of Physics and Quantum Centre of Excellence for Diamond and
Emergent Materials (QuCenDiEM), Indian Institute of Technology Madras, Chennai 600036, India}

\author{Johannes Reuther}
\email{johannes.reuther@fu-berlin.de}
\affiliation{Helmholtz-Zentrum Berlin f\"ur Materialien und Energie, Hahn-Meitner-Platz 1, 14109 Berlin, Germany}
\affiliation{Dahlem Center for Complex Quantum Systems and Fachbereich Physik, Freie Universität Berlin, 14195 Berlin, Germany\looseness=-1}
\affiliation{Department of Physics and Quantum Centre of Excellence for Diamond and
Emergent Materials (QuCenDiEM), Indian Institute of Technology Madras, Chennai 600036, India}

\author{Harald O. Jeschke}
\email{jeschke@okayama-u.ac.jp}
\affiliation{Research Institute for Interdisciplinary Science, Okayama University, Okayama 700-8530, Japan}
\affiliation{Department of Physics and Quantum Centre of Excellence for Diamond and
Emergent Materials (QuCenDiEM), Indian Institute of Technology Madras, Chennai 600036, India}

\begin{abstract} 
Quantum antiferromagnets based on the square-kagome lattice are proving to be a fertile platform for realizing nontrivial phenomena in frustrated magnetism. Recently, several decorated square-kagome compounds of the nabokoite family have been synthesized, allowing for experimental exploration of model Hamiltonians. Here, we carry out a theoretical analysis of {\nab} nabokoite using a Heisenberg Hamiltonian derived from density functional theory energy mapping. We employ classical Monte Carlo simulations to explain the two transitions experimentally observed in the low-temperature magnetization curve. Interestingly, the intermediate-field phase is also found in a purely two-dimensional model and is described by a spin liquid featuring subextensive degeneracy with a ferrimagnetic component. We show that this phase can be approximated by a checkerboard lattice in a magnetic field. Finally, we assess the effects of quantum fluctuations in zero fields using the pseudo-Majorana functional renormalization group method.
\end{abstract}

\maketitle 

\section*{Introduction}

Antiferromagnetic spin systems on highly frustrated lattices like kagome, hyperkagome or pyrochlore are known to host emergent many-body phenomena such as classical and quantum spin liquids characterized by non-trivial correlations~\cite{Broholm2020}. Besides these traditional examples of geometrically frustrated antiferromagnets, other lattices become classical or quantum spin liquids due to a combination of geometric and/or parametric frustration~\cite{Savary2017,Zhou2017}. Examples are the honeycomb lattice where a second neighbour coupling or Kitaev interactions can trigger spin liquid states~\cite{Kitaev-2006,Ferrari-2017}, the diamond lattice where competing interactions can lead to a spiral spin liquid~\cite{Bergman-2007,Gao2017,Iqbal2018,Chauhan-2023}, the maple leaf lattice with combinations of antiferromagnetic and ferromagnetic interactions~\cite{Gresista-2023,Gembe-2024,Schmoll-2024,Ghosh2023,Ghosh2024} or the trillium lattice which can host a dynamically fluctuating liquid state due to the formation of an effective tetra-trillium lattice~\cite{Zivkovic2021,Gonzalez24b}.

Here, we consider the square-kagome lattice, which has been suggested theoretically as a variant of the highly frustrated kagome lattice~\cite{Siddharthan2001}. Several theoretical studies have established that this lattice is of high intrinsic interest~\cite{Tomczak2003,Richter2009,Nakano2013,Derzhko2014} even if material realizations were only a future hope. Using exact diagonalisation and other approaches, Rousochatzakis {\it et al.} established a complex phase diagram in a magnetic field and as a function of the ratios of the exchange interactions~\cite{Rousochatzakis2013}. In the idealized case of a single antiferromagnetic nearest-neighbour coupling, a valence bond crystal ground state is expected~\cite{Astrakhantsev2021,Schmoll2022}. The theory predictions have provided ample motivation to identify material realizations of the square-kagome Hamiltonian. The first success in this respect is the synthesis of Al-atlasovite {\Alatla} by Fujihara {\it et al.} where nonmagnetic Al is replaced for the magnetic Fe of the mineral atlasovite {\atla}~\cite{Fujihala2020, Popova1987}. While the synthesis has been repeated and the highly frustrated nature of Al-atlasovite has been confirmed~\cite{Liu2022}, the nature of the Hamiltonian is still unclear. A fully synthetic square-kagome material {\skag} was realized by Yakubovich {\it et al.}~\cite{Yakubovich2021}, and even though it has an additional magnetic site, analysis of the Hamiltonian~\cite{Niggemann2023} leads to the conclusion that the decorated square-kagome lattice can support a quantum paramagnetic ground state. 

In this work, we focus on a material which may be a case where ``a spin liquid exists in a long-forgotten drawer of a museum", in the words of Broholm {\it et al.}~\cite{Broholm2020}. Nabokoite {\nab} is a mineral that was first described over three decades ago~\cite{Popova1987,Pertlik1988}. Very recently, Murtazoev {\it et al.}~\cite{Murtazoev2023} managed to synthesize not only nabokoite but also several variants like Na-nabokoite {\Nanab}, Rb-nabokoite {\Rbnab} and Cs-nabokoite {\Csnab}, establishing a whole family of compounds where fine-tuning of magnetic properties is feasible. All nabokoite variants so far feature a seventh Cu site decorating the six-site square-kagome unit cell, but it is still unclear how this impacts the magnetic behaviour. The magnetic characterization of nabokoite by Markina {\it et al.}~\cite{Markina2024_main} establishes it as a highly frustrated compound with a N{\'e}el temperature of only $T_{\rm N}=3.2$\,K compared with a Curie-Weiss temperature of $\theta_{\rm CW}=-153.6$\,K. It also features a highly unusual magnetization curve at low temperatures which exhibits two phase transitions. To date, the appropriate Hamiltonian that accurately captures the properties of nabokoite remains unidentified.

In the following, we will first establish the relevant Hamiltonian for nabokoite. It features a set of strong antiferromagnetic Heisenberg couplings, and interestingly, one of the triangle couplings of the square-kagome lattice is negligible. This calls for establishing the magnetic properties of this Hamiltonian on an effectively new lattice which has hitherto not been investigated. The first insights are obtained from classical Monte Carlo simulations. We demonstrate that the Hamiltonian can fully explain the intricate magnetization curve and two finite-temperature phase transitions ascribed to a weak inter-layer coupling. We find that the 2D regime of the Hamiltonian could be accessed experimentally by applying a finite magnetic field, which then motivates us to investigate the properties of a single decorated square-kagome layer. We show that there is subtle cooperation between the three magnetic sublattices which is responsible for inducing spin liquid behaviour. To highlight this interplay and the associated spectroscopic signatures, we propose a simplified model. Finally, we employ the pseudo-Majorana functional renormalization group approach to assess the impact of quantum fluctuations and make predictions for future inelastic neutron scattering experiments.

\section*{Results}

\subsection*{Nabokoite Hamiltonian}

\begin{figure*}[!t]
\centering
\includegraphics*[width=\textwidth]{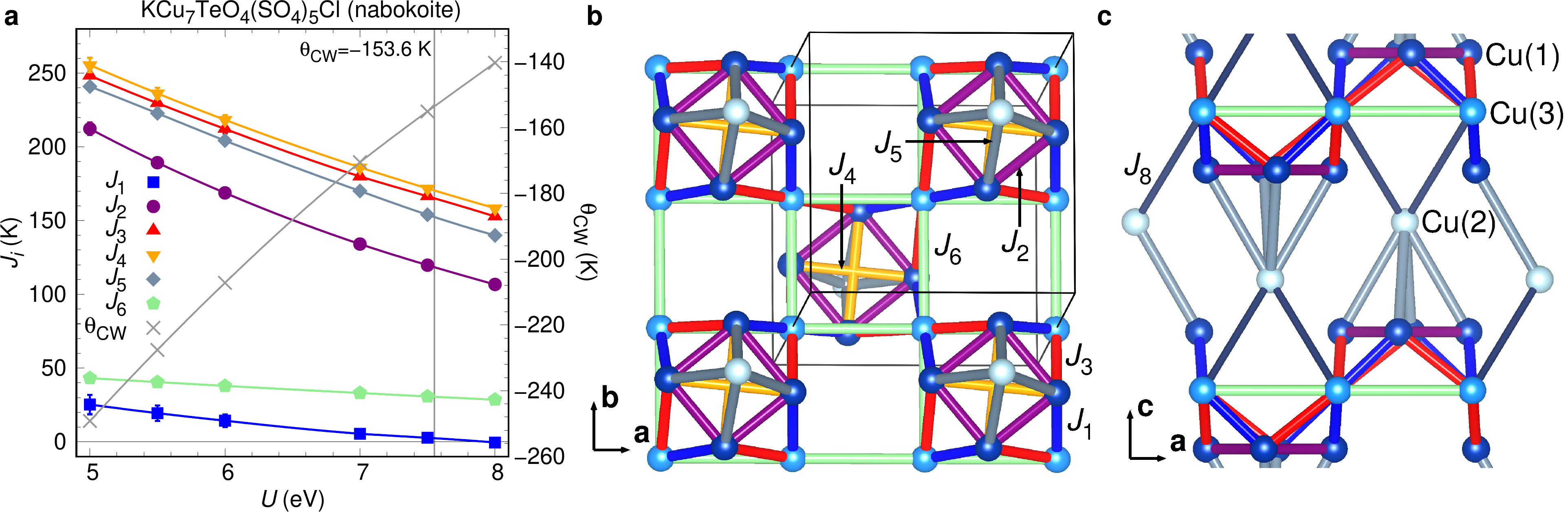}
\caption{{\bf DFT energy mapping and exchange pathways for nabokoite.} {\bf a} Values of the first six exchange interactions as a function of on-site interaction strength $U$. The vertical line marks the set of couplings that matches the experimental Curie-Weiss temperature $\theta_{\rm CW}=153.6$\,K (see the text for details). {\bf b} Exchange network defined by the six dominant exchange interactions in the decorated square-kagome plane. {\bf c} 3D connectivity established by the non-zero interlayer coupling $J_8$. Every Cu(2) is connected to four Cu(3) via $J_8$.}
\label{fig:hamiltonian}
\end{figure*}

We begin by extracting the Heisenberg Hamiltonian parameters for nabokoite using density functional theory (DFT)-based energy mapping (see Methods for technical details). This approach has been applied successfully to many highly frustrated Cu magnets~\cite{Jeschke2015, Iqbal-2015, Iida-2020, Chillal2020, Hering2022, schmoll2024_spangolite}. We base our calculations on the structure determined by Pertlik and Zemann~\cite{Pertlik1988}. We determine the parameters of the Heisenberg Hamiltonian
\begin{equation}
    H=\sum_{i<j} J_{ij} {\bf S}_i\cdot {\bf S}_j
    \label{eq:H}
\end{equation}
which we define without double-counting the exchange paths. Fig.~\ref{fig:hamiltonian}\,{\bf a} shows the result of this calculation for six values of the on-site Coulomb interaction strength $U$. We fix the energy scale of the Hamiltonian by relying on the information about the Curie-Weiss temperature contained in the experimental susceptibility $\chi(T)$ of nabokoite. While Ref.~\onlinecite{Markina2024_main} points out the difficulty in performing a linear fit of $\chi^{-1}(T)$, we apply the method proposed recently by Pohle and Jaubert~\cite{Pohle2023_main} and obtain $\theta_{\rm CW}=-153.6$\,K (see Supplementary Note 1). The exchange interactions for which the theoretical $\theta_{\rm CW}$ estimate matches this value is marked by a vertical line in Fig.~\ref{fig:hamiltonian}\,{\bf a}. The values of the six dominant couplings are listed in Table~\ref{tab:couplings}, and the network they form is illustrated in Fig.~\ref{fig:hamiltonian}\,{\bf b}. While we use the $\theta_{\rm CW}$ to obtain information about the approximate energy scales of interactions in nabokoite, we would like to point out that a rather large range of $U$ values leads to similar sets of exchange interactions (see Fig.~\ref{fig:hamiltonian}\,{\bf a}) so that the conclusions of our investigation do not strongly depend on the precise value of $\theta_{\rm CW}$. Note that the value of the on-site interaction $U=7.55$\,eV we find to describe the magnetism of nabokoite well is very much in line with typical $U$ values for Cu$^{2+}$ ions (see for example Refs.~\cite{Chillal2020,Iida-2020}). We also determine some longer range exchange couplings in the square-kagome plane but their strength is at most 5{\%} of the strongest coupling $J_4$. Furthermore, we find the interlayer coupling $J_8$ to be $J_8=-0.053~J_4$; the way this connects the square-kagome layers along $c$ is shown in Fig.~\ref{fig:hamiltonian}\,{\bf c}.
It is interesting that the nearest neighbour coupling comes out extremely small. Such a behaviour is usually due to some cancellation between contributions from different exchange paths, and finding this behaviour attests to the lack of bias in the DFT energy mapping approach. Note that a similar situation is found in \ce{K2Ni2(SO4)3}\cite{Zivkovic2021, Gonzalez24b} where the Hamiltonian has strong support due to excellent comparison with experiments.

\begin{table}
    \begin{tblr}{lrr}
        \hline
        \hline
        Exchange & in K & in J$_4$  \\
        \hline
        J$_1$ & 2(3) & 0.012 \\
        J$_2$ & 118(2) & 0.694  \\
        J$_3$  & 165(2) & 0.971 \\
        J$_4$ & 170(2) & 1.000 \\
        J$_5$ & 152(2) & 0.894 \\
        J$_6$ & 31(2) & 0.182 \\
        J$_8$ & -9(1) & -0.053 \\
        \hline
        \hline
    \end{tblr}
\caption{{\bf Exchange coupling values for nabokoite.} Couplings obtained with DFT energy mapping (see the text for details), in units of K and of the largest coupling $J_4$.}
\label{tab:couplings}
\end{table}

\subsection*{Magnetization process}

\begin{figure*}[!t]
\centering
\includegraphics*[width=\textwidth]{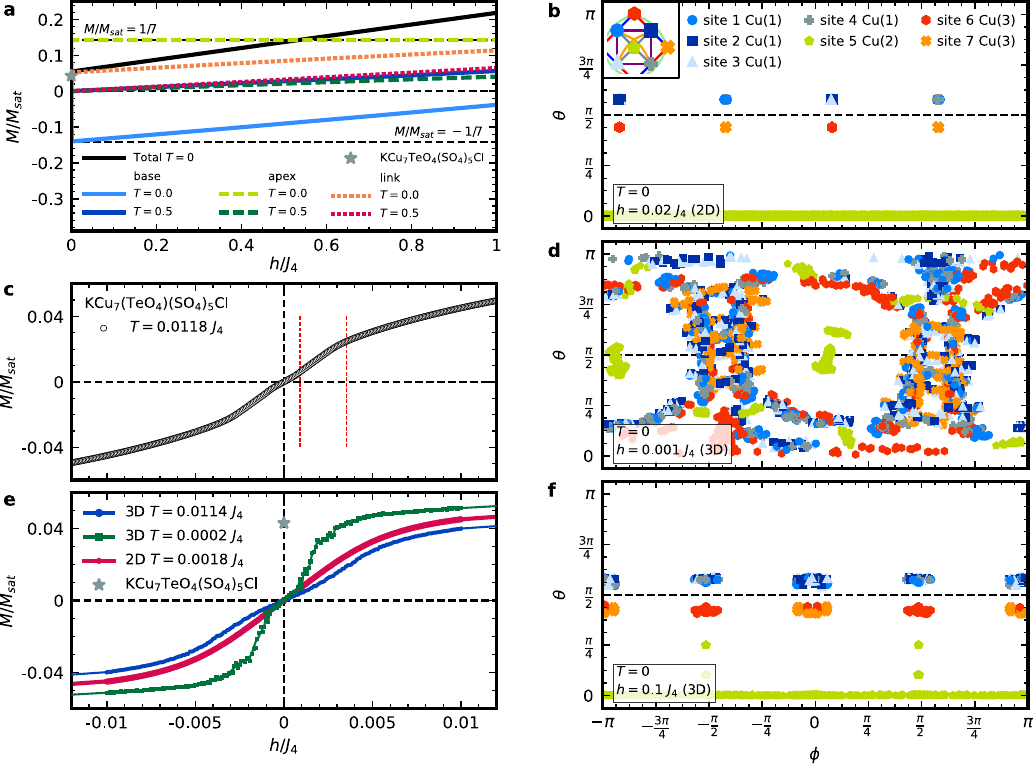}
\caption{{\bf Magnetization curves.} {\bf a} Magnetization process for the 3 sublattices on the 2D model at two different temperatures, $T=0$ and $T=0.5J_4$ indicated by two different colours (light and dark colours, respectively). The black line shows the total magnetization at $T=0$. The star corresponds to the {\nab} data from Ref.~\cite{Markina2024_main}. {\bf b} Configuration in polar coordinates for $h=0.02\,J_4$ at $T=0$ in the 2D model. Each point indicates a spin, and colours denote the site within the unit cell: Cu(1) (pyramid base), Cu(2) (apex), Cu(3) (link). {\bf c} Magnetization curve from Ref.~\cite{Markina2024_main} corresponding to $T= 0.0118\,J_4$ ($T=2$\,K). {\bf d} Spin configuration at $T= 0$ for the 3D model where the magnetic fields is $h= 0.001\,J_4$. {\bf e} Magnetization curve for the 2D and 3D dimensional models at different temperatures, where the grey star corresponds to the $h\to0$ extrapolation made from higher fields in Ref.~\cite{Markina2024_main}. {\bf f} Spin configuration at $T= 0$ for the 3D model for $h=0.1\,J_4$.}
\label{fig:mag}
\end{figure*}

The experimentally measured magnetization curve for nabokoite {\nab} exhibits two transitions at finite temperatures. In the data taken from Ref.~\cite{Markina2024_main} (Fig.~\ref{fig:mag}\,{\bf c}), there is an initial small slope at small fields which then grows and decreases again, defining three different regimes at finite temperatures. To evaluate this theoretically, we perform classical Monte Carlo (cMC) calculations on two different lattices. As a first approach, we disregard the inter-layer coupling $J_8$ which is ferromagnetic and only $\sim 5\%$ of the largest coupling. The model then consists of a single layer and we, henceforth, refer to it as the two-dimensional (2D) model. In contrast, when considering the full Hamiltonian with inter-layer coupling we will speak of the three-dimensional (3D) model. While continuous spin symmetries cannot be broken at finite temperatures in two-dimensional lattices due to Mermin and Wagner's theorem~\cite{Mermin66}, it is not uncommon to find finite-temperature phase transitions associated with broken discrete symmetries~\cite{Chandra90, Capriotti04, Grison20, Gonzalez24a}. In this case, we observe a small peak at very low temperatures that does not scale with system size and is therefore not associated with a phase transition (see Supplementary Figure 3).

The magnetization process for the 2D model under a magnetic field $h$ is shown in Fig.\,\ref{fig:mag}\,{\bf e}. The sublattice magnetizations (corresponding to each symmetry inequivalent site) are shown in Fig.~\ref{fig:mag}\,{\bf a} for two different temperatures $T=0$ and $T=0.503\,J_4$ in light and dark colour, respectively. The total magnetization at $T=0$, shown by the black line, hints that the ground state has a finite magnetization in the $h\to 0$ limit, indicating a ferrimagnetic behaviour for the 2D model. From the sublattice magnetization, we can see that the ground state has all decorating spins at the pyramid apices aligned ferromagnetically, amounting to $1/7$ of the total magnetization. We could have expected then that the total non-zero magnetization originates from these spins. However, the contribution of the pyramids to the total magnetization at low temperatures is very small compared to the total magnetization itself (black curve), since the light-blue and light-green lines almost cancel each other at low fields and temperatures. This happens because the spins in the base of the pyramid oppose and cancel the pyramid's ferrimagnetic magnetization almost perfectly. It is then the link sites connecting the pyramids that are responsible for the total non-zero magnetization, as can be seen by the low-field agreement between the orange dotted line and the black curve.

The high slope observed experimentally for the intermediate phase at finite temperature can be related to the non-zero value of $M$ at $h=0$ and $T=0$, for which the connecting sites are responsible. In other words, the non-zero value of $M$ at $T=0$ softens into a sharp increase of $M$ at small fields as the temperature increases (see red curve in Fig.~\ref{fig:mag}\,{\bf e}). On the other hand, the smaller slope at higher fields comes from the spins at the base of the pyramid. These spins, which at small fields are pointing against the apex spins and the magnetic field, start aligning slowly with the magnetic field as can be seen by comparing the black curve to the light-blue one. Experimentally, the extrapolated magnetization from the high-field slope to $T=0$ is $M=0.043\,M_{\rm sat}$ (where $M_{\rm sat}$ is the magnetization of the fully saturated state)~\cite{Markina2024_main}. This extrapolation is performed at higher fields than the ones shown in Fig.~\ref{fig:mag}\,{\bf c}, where the behaviour is linear~\cite{Markina2024_main}. In our case, extrapolating the total magnetization to $h=0$ and $T=0$ leads to $M=0.055\,M_{\rm sat}$ which is surprisingly close to the experimental value of $0.043\,M_{\rm sat}$, denoted by a star in Fig.~\ref{fig:mag}\,{\bf a}. However, the precision could be accidental since there is {\it a priori} no reason for the classical $S=\infty$ simulations to quantitatively reproduce the results for a $S=1/2$ compound at zero field and temperature. We note that our calculations for the 2D model miss capturing the low-field phase with the small slope of the magnetization curve -- we will subsequently show that the inclusion of the inter-layer coupling is crucial for capturing this subtle phase transition.

We first analyze the spin configuration at low temperatures and fields, which is shown in Fig.~\ref{fig:mag}\,{\bf b}. There, we plot a spin configuration at $h=0.02\,J_4$ and $T=0$ using polar coordinates of the $|\mathbf{S}|=1$ spins, where $\theta$ is the angle with the $z$ axis and $\phi$ is the angle in the $xy$-plane. The green dots around $\theta=0$ indicate that all the apices of the pyramids are pointing nearly in the $z$ direction. Then, the pyramid's base presents spins in 4 directions slightly canted down from the $xy$ plane as shown by the values of $\theta$ higher than $\pi/2$. This locking of the spins in four directions is what causes the peak we observe in the specific heat at very low temperatures (Supplementary Figure 3). We can also observe that spins connected by the diagonal coupling $J_4$ are mixed (blue and grey on the one hand, and light blue and dark blue on the other), meaning that they are exchanged in different unit cells. The reason will become clear from the ground state of the single pyramid discussed below. Finally, the sites linking the pyramids (red and orange) are slightly canted towards the magnetic field, in the same direction as the pyramid apices. These spins are also aligned in terms of $\phi$ with the base of the pyramids and can be separated into two groups (orange and red symbols do not mix).

The effect of the interlayer ferromagnetic coupling $J_8$, even though being only $\sim 5 \%$ of the largest coupling, is far from trivial. Firstly, the 3D model exhibits a finite-temperature phase transition for $h=0$ at $T_{\rm{c}} = 0.0095\,J_4$ (see Supplementary Figure 7), which is the same order of magnitude as the experimental value $0.019\,J_4$~\cite{Markina2024_main}. Even though the states above the phase transition are similar to those of the 2D model, we find that the ground states of both models completely differ at $h=0$. This can be seen in the $(\phi,\theta)$ plot in Fig.~\ref{fig:mag}\,{\bf d}, where a configuration for $T=0$ and $h=0.001\,J_4$ is shown. The small field does not change the state but makes the pattern easier to see (see Supplementary Notes 3 and 4 for more details on the spin configurations).

Since the 2D model failed to capture the low-field phase transition observed experimentally, it is to be expected that the magnetization process of the 3D Hamiltonian will exhibit a new phase transition from the low-field regime where the interlayer coupling $J_8$ selects the ground state to a high-field regime where the system neglects this small coupling. This becomes clear in the magnetization curve of Figure~\ref{fig:mag}\,{\bf c}, which shows the experimental results for {\nab} at $T=2$\,K taken from Ref.~\cite{Markina2024_main} along with the two phase transitions they determined (shown in red dashed lines). The value of the magnetization $M$ is shown as a fraction of the saturation magnetization $M_{\rm sat}$, and the magnetic field is converted to units of $J_4$. Our cMC results are presented in Figure~\ref{fig:mag}\,{\bf e}, where the blue magnetization curve is calculated for the 3D Hamiltonian at the temperature of the experiments on {\nab} and exhibits an impressive agreement (especially considering that we are comparing $S=1/2$ and $S=\infty$). 

Most importantly, our calculations capture the key aspect of three different regimes separated by two transitions. At very low magnetic fields, there is an initial regime in which the magnetization grows slowly. This is succeeded by an increase in the slope (susceptibility) where the magnetization grows fast, and then it flattens again reaching a similar situation to the one observed in the two-dimensional model (red curve). These features of the 3D model are enhanced at low temperatures (green curve), where it becomes evident that the low-temperature regime is particular to the 3D model (red versus green curves). As anticipated, this is caused by the difference in the ground state: whereas in 2D there is a finite value of the total magnetization that leads to a ferrimagnetic behaviour, in 3D this does not happen. The good agreement between our 3D model calculations in Fig.~\ref{fig:mag}\,{\bf e} at $T=0.0114~J_4$ (blue curve) and the experimentally measured magnetization curve in Fig.~\ref{fig:mag}\,{\bf c} allows us to conjecture that {\nab} does not have a ferrimagnetic behaviour at $T=0$ and $h=0$; a small magnetic field is needed to enter the ferromagnetic phase.

These results not only validate the Hamiltonian we derived using DFT-based energy mapping, which provides the small ferromagnetic interlayer coupling that is needed to reproduce the experimental results; it also allows us to make some relevant predictions for future experiments. We find here that the two-dimensional regime can be accessed at finite fields in the 3D model; this implies that the spin liquid properties of the 2D model we discuss below are experimentally accessible. This is further supported if we look at the spin configurations of the states in the low-field and high-field regimes, Figures~\ref{fig:mag}\,{\bf d} and {\bf f}, respectively. The low-field configuration close to the ground state at zero field has the pyramid apex spin pointing mostly in-plane while the rest of the spins form complicated patterns around them. Eventually, when the magnetic field is increased, we observe a transition to the states of Fig.~\ref{fig:mag}\,{\bf f} which resemble those of Fig.~\ref{fig:mag}~{\bf b} for the 2D Hamiltonian. We show this effect for $h = 0.1\,J_4$ which corresponds to about $25$\,T, but similar states can be accessed at lower fields.

\subsection*{Spin-liquid features in the 2D model}

We proceed to analyze the properties of the 2D model for nabokoite {\nab} in more depth. Even though the ground state differs from that of the 3D model, at finite temperatures the states are similar. Furthermore, we showed in the previous section that the ground state of the 2D model can be reached in the 3D model -- in other words, in an experiment -- under a small magnetic field. Therefore the 2D model is not only interesting theoretically, but also experimentally accessible. With cMC the ground-state energy can be obtained as a continuation of the cool-down protocol till $T=0$. By doing so, we find that lattices with even $L$ reach lower ground-state energies compared to those with odd $L$, indicating that the latter are frustrated by the periodic boundary conditions (see Supplementary Figure 3). It is also important to note that lattices with even $L$ are not frustrated by the boundary conditions because they all reach the same ground-state energy, given by the one on the $L=2$ cluster. We will show below that this property is not related to a doubling of the magnetic unit cell with respect to the structural unit cell, but with a fluctuation mechanism that necessitates the even periodicity of the lattice.

\begin{figure*}[!t]
\centering
\includegraphics*[width=\textwidth]{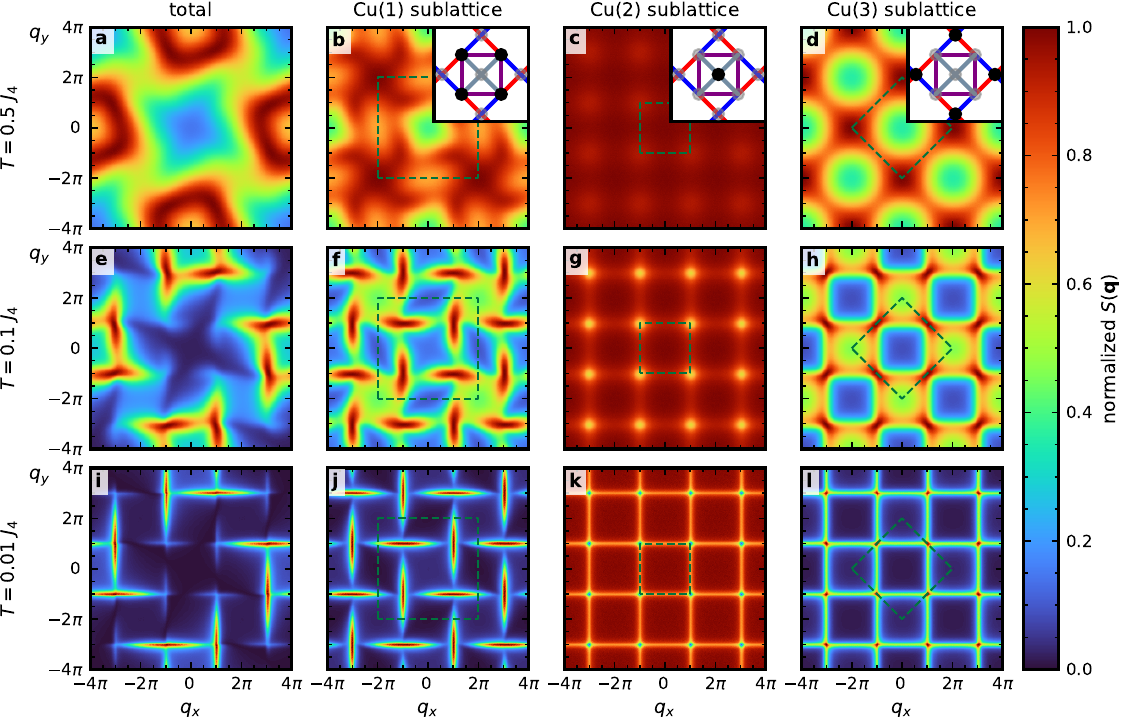}
\caption{{\bf Spin structure factors of the 2D Hamiltonian}. $S(\mathbf{q})$ for three different temperatures $T=0.5\,J_4$ ({\bf a} to {\bf d}), $T=0.1\,J_4$ ({\bf e} to {\bf h}), and $T=0.01\,J_4$ ({\bf i} to {\bf l}). The first column ({\bf a}, {\bf e}, {\bf i}) shows the result for the entire lattice, and the other three columns give sublattice results: column two ({\bf b}, {\bf f}, {\bf j}) for the pyramid base, column three ({\bf c}, {\bf g}, {\bf k}) for the pyramid apex, and column four ({\bf d}, {\bf h}, {\bf l}) for the linking site, as shown in the insets of {\bf b}, {\bf c} and {\bf d}. The Brillouin zones corresponding to the sublattices are denoted by dashed green squares on each panel. All spin structure factors are normalized.}
\label{fig:ssf_cmc}
\end{figure*}

\begin{figure*}[!t]
\centering
\includegraphics*[width=\textwidth]{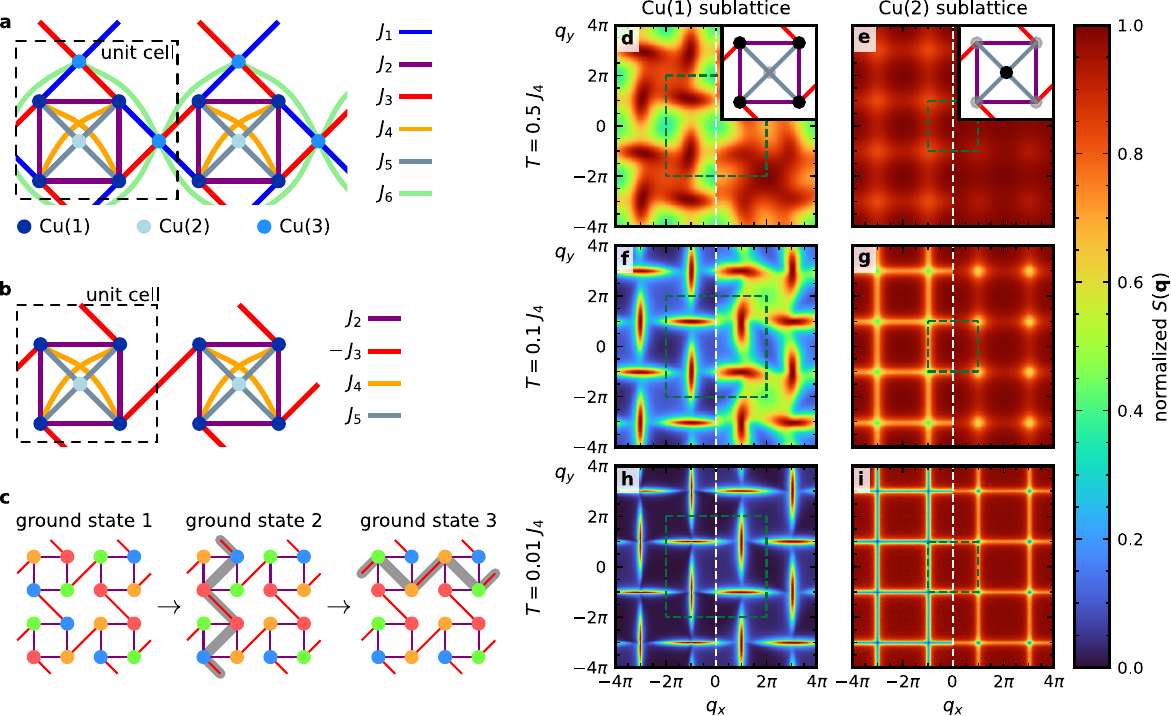}
\caption{{\bf Effective model for the ferrimagnetic spin liquid}. {\bf a} 2D model planar lattice without spatial distortions. {\bf b} Effective lattice for the spin liquid model where Cu(3) sites as well as $J_1$ and $J_6$ exchange interactions have been removed. The sign of $J_3$ is flipped to account for the absence of Cu(3) sites. {\bf c} Ground-state configurations for the effective model, where each colour represents a different spin direction. Ground states 1 and 2 are connected by exchanging the spins along the grey path. A second such exchange connects ground states 2 and 3. {\bf d} to {\bf l} Comparison of the spin structure factor of the effective model (left side of each panel) with the full two-dimensional model (right side) for three different temperatures and the two relevant sublattices (indicated in the insets of {\bf d} and {\bf e})}
\label{fig:latt3}
\end{figure*}

To analyze the spin correlations we calculate the equal-time spin structure factor, defined by
\begin{equation}
    S(\mathbf{q}) = \frac{1}{N} \sum_{i,j} \ \langle \mathbf{S}_i \cdot \mathbf{S}_j \rangle \ \rm{e}^{\rm{i}\mathbf{q}\cdot \mathbf{r}_{ij}}
\end{equation}
where $\mathbf{r}_{ij} = \mathbf{r}_i -\mathbf{r}_j$ is the distance vector between spins at positions $\mathbf{r}_i$ and $\mathbf{r}_j$. Apart from calculating the spin structure factor from the whole lattice, it is informative to also calculate it for the different sublattices in the system, especially when there are symmetry-inequivalent sites. Therefore, we also calculate $S(\mathbf{q})$ for the three sublattices formed by the three different types of sites (see Fig.~\ref{fig:hamiltonian}). To analyse the spin structure factor it is convenient to simplify the structure by setting all the spins in the same plane and eliminating the twisting distortions of the pyramids (see Fig.~\ref{fig:latt3}\,{\bf a}). The results for three different temperatures are shown in Fig.~\ref{fig:ssf_cmc}:  $T=0.5\,J_4$ in Fig.~\ref{fig:ssf_cmc}\,{\bf a} to {\bf d}, $T=0.1\,J_4$ in Fig.~\ref{fig:ssf_cmc}\,{\bf e} to {\bf h}, and $T=0.01\,J_4$ in Fig.~\ref{fig:ssf_cmc}\,{\bf i} to {\bf l}. The spin structure factor for the whole lattice is shown in Fig.~\ref{fig:ssf_cmc}\,{\bf a}, {\bf e} and {\bf i}, while $S(\mathbf{q})$ for the three sublattices is shown in the three subsequent columns of Fig.~\ref{fig:ssf_cmc}. Insets in Fig.~\ref{fig:ssf_cmc}\,{\bf b}, {\bf c} and {\bf d} indicate the shown sublattice by black circles. Dashed lines show the first Brillouin zones for the sublattices.

At intermediate temperatures ($T=0.5\,J_4$), the total spin structure factor exhibits a continuum of peaks along rings centred around $\mathbf{q}=(\pm 4\pi, 0)$ and $(0,\pm 4\pi)$. Even though these are not perfectly homogeneous, the variation of the intensity along the ring is about $5\%$ of the maximum, indicating the presence of a spiral liquid-like regime~\cite{Iqbal2018}. The chirality (or the breaking of the reflection symmetries $q_x,q_y \to -q_x,q_y$ and $q_x,q_y \to q_x,-q_y$) is given by the difference between $J_1$ and $J_3$, the two bonds that connect the pyramids to the Cu(3) spins. This chirality is also reflected in the structure factor of the Cu(1) sublattice (Fig.~\ref{fig:ssf_cmc}\,{\bf b}), in which the peaks are centred in the corners of the first Brillouin zone of the pyramid base sublattice, showing a tendency towards antiparallel nearest neighbours or N\'eel order. However, these peaks spread in four directions resembling a twisted windmill. The spin structure factor of the Cu(2) sites at the pyramid apex (Fig.~\ref{fig:ssf_cmc}\,{\bf c}) is practically featureless at $T=0.5\,J_4$, indicating that it behaves mostly as an uncorrelated paramagnet. Finally, the Cu(3) spins that link the pyramids are shown in Fig.~\ref{fig:ssf_cmc}\,{\bf d} and also exhibit peaks in the corners of their respective Brillouin zone corresponding to a N\'eel ordering tendency.

Lowering the temperature to $T=0.1\,J_4$ and $T=0.01\,J_4$ leads to interesting changes in the spin structure factor (Figs.~\ref{fig:ssf_cmc}\,{\bf e} to {\bf h} and {\bf i} to {\bf l}, respectively). In the total $S(\mathbf{q})$, the spiral rings start to decompose and peaks start to form at $\mathbf{q}=(\pm \pi,\pm 3\pi)$ and $(\pm3\pi,\pm \pi)$. At the lowest temperature, these peaks have the particular characteristic of forming needles along very specific Cartesian directions. The needles indicate that the system has the freedom to fluctuate in specific linear directions, and are also evident in the spin structure factor of the Cu(1) sublattice, formed by the spins of the pyramid base. For this sublattice (Fig.~\ref{fig:ssf_cmc}\,{\bf f}), the windmills start to spread out and the peaks disappear from the corners of the Brillouin zone, leading to the needles observed in Fig.~\ref{fig:ssf_cmc}\,{\bf j}. Below, we will use an effective model to show that these needles are related to a subextensive degeneracy given by zero modes which involve exchanging all base spins across a zig-zag line that crosses the system in one Cartesian direction. A similar behaviour occurs in the breathing pyrochlore lattice, where planes can be flipped in a certain range of parameters leading to square rings in the spin structure factor~\cite{Benton15}.

Continuing with the pyramid apex site (Fig.~\ref{fig:ssf_cmc}\,{\bf g} and {\bf k}), the spin structure factor presents an unusual behaviour. When the temperature is lowered, the weak high-temperature maximum around $\mathbf{q}=(0,0)$ (Fig.~\ref{fig:ssf_cmc}\,{\bf c}) spreads and evenly distributes over the entire Brillouin zone except for the wave-vectors belonging to the zone boundary which lose intensity. At the lowest temperature of $T=0.01\,J_4$ (Fig.~\ref{fig:ssf_cmc}\,{\bf k}), there is a constant intensity with a square shape outlined by $S(\mathbf{q})$ minima. This property will also be analysed below using the effective model and is related to the freedom of the apex spins to fluctuate because they are only connected to the system by $J_5$. Finally, the Cu(3) sublattice (Fig.~\ref{fig:ssf_cmc}\,{\bf h} and {\bf l}) exhibits a shift of the peaks from the corners of the Brillouin zone towards the mid-point of the Brillouin zone edge. Peaks at this high symmetry point are typically associated with stripe-like patterns. In this case, both stripe directions coexist and the spin structure factor at $T=0.01\,J_4$ (Fig.~\ref{fig:ssf_cmc}\,{\bf l}) exhibits intensity along the lines that connect the high symmetry points, indicating again that the system can fluctuate in specific directions. 

\subsection*{Proximity to an exact classical spin liquid}

We have shown that the two-dimensional model for the nabokoite compound {\nab} exhibits interesting features in the classical limit, such as needles in the low-temperature spin structure factor. To understand the physical processes better, we introduce a simplified effective model that retains the key features of the 2D model. The first and obvious step is to neglect $J_1$. Even though it is the nearest-neighbour interaction in the material, the DFT Hamiltonian reveals that its value is only about $1\%$ of the largest coupling $J_4$. The next coupling that can be disregarded is $J_6$, which connects the Cu(3) spins into a square lattice (see Fig.~\ref{fig:latt3}\,{\bf a}). It takes values of about $18\%$ of $J_4$, which is not so small. However, since this coupling tends to order the Cu(3) sublattice into a N\'eel state and the corners of the Brillouin zone present the lowest signal in $S(\mathbf{q})$ at low temperatures (see Fig.~\ref{fig:ssf_cmc}\,{\bf l}), we can argue that it does not play a key role in the low-temperature physics. Our results below will also justify the simplification \textit{a posteriori}. Eliminating $J_1$ and $J_6$ leads to a lattice of pyramids coupled to nearest-neighbour pyramids only by a combination of two $J_3$ couplings and a Cu(3) spin. The bonds $J_3-\mathbf{S}_3-J_3$ connecting neighbouring pyramids are unfrustrated and can be replaced simply by a ferromagnetic coupling $-J_3$. The resulting effective lattice which retains only four bonds and two types of spins is shown in Fig.~\ref{fig:latt3}\,{\bf b}. The unit cell is reduced from 7 to 5 spins. It is important to note that this procedure of replacement of two antiferromagnetic couplings by a ferromagnetic coupling $-J_3$ strictly only holds for the ground state and that at non-zero $T$, the effective coupling $-J_3$ will vary with temperature, leading to only a qualitative mapping. While not carried out in the present work, it would be worthwhile to employ the decoration-iteration method~\cite{Strecka-2015} to integrate out such unfrustrated degrees of freedom as has been done previously to map this class of spin models on the square kagome lattice onto the checkerboard lattice~\cite{Pohle2016}.

The specific heat calculations on this model do not exhibit any signatures of a phase transition at finite temperatures, indicating that no discrete symmetries are broken. Furthermore, the specific heat does not exhibit any peak at low temperatures in contrast to the full 2D model. The reason is that the base spins are not locked into four specific directions in this case (see Supplementary Figure 6). As the Cu(3) spins are missing from this model, we forego the comparison of the total spin structure factor to the full model and instead focus on  $S(\mathbf{q})$ for Cu(1) and Cu(2) sublattices. The results are shown in Figs.~\ref{fig:latt3}\,{\bf d} to {\bf i}, with the effective model on the left and the full model on the right of each panel (full model results are the same as in Fig.~\ref{fig:ssf_cmc}). We use the same three temperatures but expect differences only at temperatures below the two eliminated couplings. However, Fig.~\ref{fig:latt3} shows that the effective model reproduces the spin structure factor of the {\nab} Hamiltonian well for lower temperatures. Most importantly, at $T=0.01\,J_4$ (Figs.~\ref{fig:latt3}\,{\bf h} and {\bf i}), the effective model reproduces the same needle structure for the Cu(1) sublattice that is observed in the original model. The features of the Cu(2) sublattice are also reproduced. This is counterintuitive in the sense that the eliminated couplings represent the lowest energy scales of the system and we would expect their influence to be stronger at low temperatures. However, we find that the fundamental physical mechanisms occurring in the original Hamiltonian can be studied from, and are well represented in, the simpler effective model. 

Let us first focus on a single pyramid from Fig.~\ref{fig:latt3}\,{\bf b} and ask how the energy can be minimized in such a five-spin system. Even though every spin in the pyramid interacts with every other, due to the difference in the couplings $J_2 \neq J_4$, the corresponding five-spin Hamiltonian cannot be rewritten as a complete square. We first focus on the four Cu(1) spins connected in a square with $J_2$ on the sides and $J_4$ diagonals, and disregard the apical spin and the $J_5$ coupling. This square has two possible solutions depending on the ratio of antiferromagnetic $J_2$ and $J_4$ couplings. For $J_4 < J_2$, the spins minimize the energy by forming a collinear N\'eel state on the square. In contrast, if $J_4 > J_2$ as in the present case, the energy is minimized by a coplanar state in which each diagonal has opposing spins, but the angle between spins of the two diagonals is completely free. These are the same states that occur on a square lattice with next-nearest neighbour interactions, where entropy favours stripe phases for the $J_4 > 0.5 J_2$ case that lead to an emergent discrete $\mathcal{Z}_2$ symmetry~\cite{Chandra90, Weber03}. However, in the present case, the angle between spins in different diagonals is completely free. Let us then assume that the coplanar state on the square is in the $xy$ plane, and let us couple the centre spin pointing in the $z$ direction to the square by turning on $J_5$. For $J_5\to 0$, the state on the square will remain coplanar. When $J_5$ increases, the spins on the square will start canting outside the plane, leading to a total net ferrimagnetic moment. As $J_5 \to \infty$, the ground state has the centre spin pointing up and all the rest pointing down (see Supplementary Note 6 for more details).

Once we know the ground state of a pyramid, the ferromagnetic coupling $-J_3$ only copies the spin from one corner of a pyramid to the corner of another pyramid. If it is possible to cover the whole lattice this way (in other words, if $-J_3$ is not a frustrating interaction), this serves as a ground-state configuration for the effective model. As stated above, the four spins belonging to a pyramid base group into two diagonal sets. In each diagonal, the spins point approximately in opposite directions (with deviations owing only to the shared canting angle). In the following analysis, we use the states corresponding to the full 2D model, where the angle between spins in opposing diagonals is nearly $90^\circ$, as shown already in Fig.~\ref{fig:mag}\,{\bf b}. We then assign four colours to these four spins: red, blue, green, and orange, corresponding to the four clouds of base spins shown in Fig.~\ref{fig:mag}\,{\bf b}. Then, we attempt to colour a $2\times 2$ lattice, as shown in Fig.~\ref{fig:latt3}\,{\bf c}: to this effect, we distribute the four colours in one of the squares and then copy the colours to other squares using the red bonds. Such a tessellation can be viewed as a constrained version~\cite{Gomez-2024} of the canonical four-colouring problem~\cite{Kondev-1995}. It is manifest that the colours of one square do not fix a ground state, and additional (arbitrary) fixing is needed. However, a ground state can be formed, indicating that $J_3$ is not a frustrating exchange interaction and that the ground state in the whole lattice consists of combinations of ground states of single pyramids.  

Furthermore, there is a way to flip spins and visit other ground states. For example, one can go from ground state 1 to 2 by exchanging red and blue spins along a vertical path consisting of diagonals and $J_3$ bonds, as shown by the grey line in Fig.~\ref{fig:latt3}\,{\bf c}. Then we can go to a different ground state by flipping a horizontal line of diagonals and red bonds. This process can be applied to any vertical or horizontal path consisting of only two colours of spins, giving rise to a subextensive degeneracy in larger lattices due to the $2L$ flippable line loops on a $L\times L$ lattice. In effect, the aforementioned constraint is so strong that choosing a pair of colours along one diagonal of a square fixes the corresponding diagonals for all squares in the same column (or row). We can represent this constraint by assigning an Ising variable (or arrow) to each column and each row. As a result, the ground states can be mapped onto the ground-state manifold of the Rys F-model in the limit $e_1=e_2=e_3=e_4 \ll 0$ and $e_5=e_6=0$, following the standard notation in Refs.~\cite{Lieb-1967,Rys-1963}. Further interactions would lift the degeneracy of this subextensive manifold, ultimately yielding a fourfold-degenerate ground state~\cite{Kondakor-2023}. 

The configurations in Fig.~\ref{fig:latt3}\,{\bf c} also explain the need for an even number of unit cells (above we found that the ground-state energy in the full Hamiltonian is only achieved for even values of $L$): a lattice with odd $L$ cannot be coloured properly, indicating the presence of frustration. Note that in all these ground states, the pyramid apex sites point in the same direction, forming a ferromagnetic state and giving the system a total net ferrimagnetic moment. However, only at very small temperatures do the centre spins align ferromagnetically. Therefore, we have shown the existence of a spin liquid with a non-zero total magnetic moment at $T=0$: a ferrimagnetic spin liquid, characterized by needles in the spin structure factor. It is important to note that the constrained four-colouring solution is valid for both the 2D and effective models, even though in the latter they represent only a subset of the ground-state manifold. As mentioned above, this is because the angles are not locked at $~90^\circ$ in the effective model and the link spins are missing. This causes spin directions to only repeat for spins connected through zig-zag lines (shown by the grey lines, for example), while spins on independent zig-zag lines are completely independent (see Supplementary Figure 6 for more details).

The effective model can be further simplified to obtain a well-known lattice. It can be thought of as a lattice of corner-sharing pyramids by rotating each pyramid counter-clockwise by $\pi/4$ and merging the spins connected by $-J_3$ into one. If we forget for a moment about the pyramid apex spins and about $J_5$ that connects them to the lattice, the result of the rotation is the checkerboard lattice. This lattice has been thoroughly studied in the classical limit and is known to harbour a classical spin liquid~\cite{Davier23}. In this case, however, the spin configuration is non-coplanar due to the presence of the spins at the centre of the squares. In our analysis of the ground state, these spins are all pointing in the same direction, such that they can be thought of as a uniform magnetic field. Therefore, the resulting Hamiltonian turns into a checkerboard lattice in a magnetic field. We emphasize though that this is only valid for the ground state and, as soon as $T\neq 0$, the spins at the top of the pyramids cease to act as a uniform magnetic field as evidenced by the mostly constant intensity in the spin structure factor of Fig.~\ref{fig:ssf_cmc}\,{\bf k} for $T=0.01\,J_4$.

\subsection*{Effect of quantum fluctuations}

The magnetic moments of the Cu$^{2+}$ ions in nabokoite {\nab} are $S=1/2$ spins, which behave quantum-mechanically at low temperatures. So far, we have only treated the DFT Hamiltonian classically and obtained some qualitative and quantitative agreement with the experimentally measured magnetization curve; this gave us important insight into the low-temperature mechanisms that drive the classical spin liquid state. However, analysing the Hamiltonian in a quantum mechanical formalism is essential to check how the physical behaviour changes. There are not many methods that can treat highly frustrated quantum systems in two dimensions and at finite temperatures. In this case, we resort to the pseudo-Majorana functional renormalization group (PMFRG) method, which relies on re-writing the spin-spin interactions in terms of Majorana operators~\cite{Niggemann21_main, Muller24}. In particular, we use the recently developed temperature flow scheme~\cite{Schneider24_main}, where the renormalization group flow equations can be written in terms of the physical temperature $T$. In this sense, we depart from a trivial known solution at $T=\infty$ and solve the flow equations as the temperature decreases to obtain the spin-spin correlations at different temperatures. The PMFRG method has proven to be reliable in two- and three-dimensional systems, obtaining critical exponents and critical temperatures in quantitative agreement with quantum Monte Carlo calculations, and qualitative signals of pinch-point singularities, among others~\cite{Niggemann22_main, Niggemann23}.

\begin{figure}[!t]
\centering
\includegraphics*[width=0.9\columnwidth]{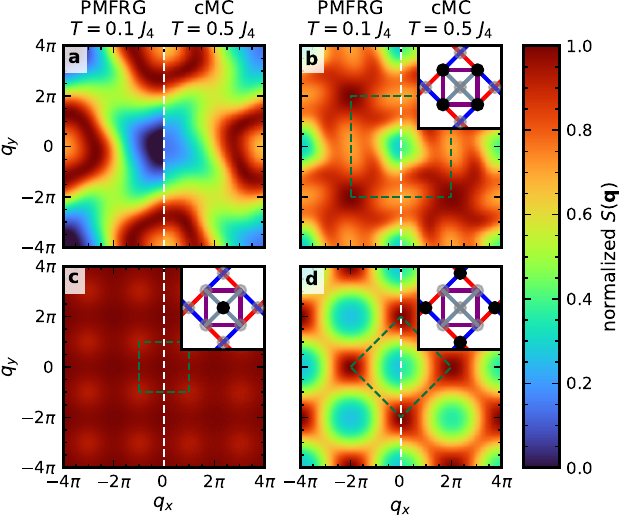}
\caption{{\bf Quantum versus classical spin structure factors.} Quantum spin structure factor calculated with PMFRG at $T=0.1\,J_4$ (left side of each panel) compared with the cMC results at $T=0.5\,J_4$. {\bf a} Total spin structure factor, {\bf b} pyramid base, {\bf c} pyramid apex, {\bf d} linking site. The sublattices are indicated by the insets.}
\label{fig:pmfrg2}
\end{figure}

In Fig.~\ref{fig:pmfrg2}, we show the calculated spin structure factor at low temperatures, $T=0.1\,J_4$, for the whole lattice and the three different sublattices of the 2D model (left side of each panel). We compare the result with the cMC calculations at $T=0.5\,J_4$, shown on the right side of each panel. We find a good agreement between quantum and classical results at higher temperatures (the cMC calculations at $T=0.1\,J_4$ look completely different, see Fig.~\ref{fig:ssf_cmc}\,{\bf e}). Such a phenomenon has been referred to as quantum-to-classical correspondence~\cite{Kulagin13, Huang16, Wang20, Gonzalez24b}, and it has been observed in the Heisenberg model on several two- and three-dimensional lattices. However, it should be pointed out that this correspondence holds only on the level of static correlations and not for the full dynamical structure factor (or time-dependent correlations). A theoretical description of why this happens has been recently given in terms of perturbation theory, where the correspondence breaks at fourth order in $J/T$~\cite{Schneider24b}. Furthermore, partial diagrammatic cancellations give rise to a good agreement even at moderately low temperatures. Focusing only on the quantum results, our calculations indicate that the spiral liquid rings remain visible down to comparatively lower temperatures compared to the classical case, such that these features can be observed experimentally in a wide range of temperatures. Furthermore, the Cu(2) sublattice of apical spins shows an almost flat $S(\mathbf{q})$, indicating that these spins remain roughly paramagnetic down to low temperatures. Thus, they can be very susceptible to small magnetic fields. In contrast to known partially disordered phases~\cite{Gonzalez19, Seifert19, Blesio23}, in this case, two different sublattices are disordered. However, one of them is correlated into a spiral-liquid-like phase, and the other behaves as an uncorrelated paramagnet. Unfortunately, $T=0$ is unreachable within PMFRG and we cannot explore the quantum ground state to check whether the classical spin liquid state from cMC has a quantum analogue. However, the experiments in {\nab} show a phase transition at finite temperatures that is probably triggered by inter-layer interactions (as we showed for the classical 3D model), such that the ground-state physics of the isolated layers is likely unreachable. Nonetheless, we showed above that the interesting spin liquid with needle-like features can be accessed experimentally at finite fields and it is not confined to the ground state in 2D and $h=0$.

\subsection*{Predictions for neutron scattering experiments}

\begin{figure*}[!t]
\centering
\includegraphics*[width=\textwidth]{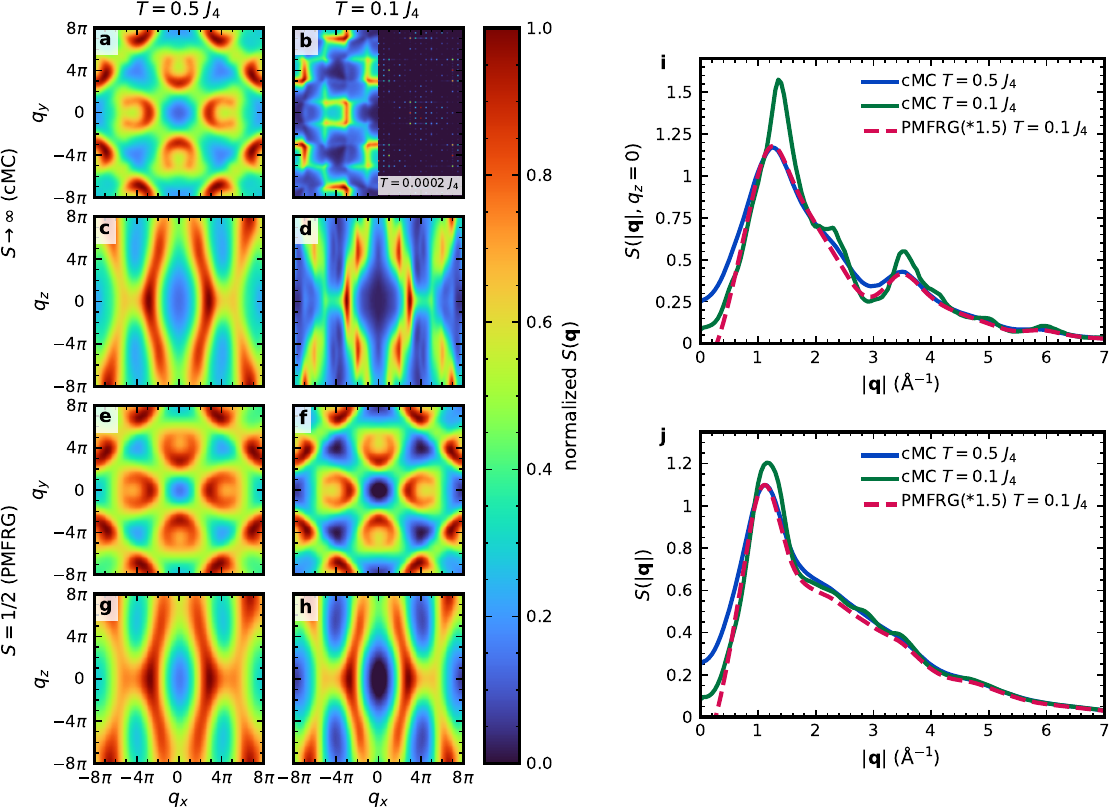}
\caption{{\bf Prediction for neutron scattering experiments}. ({\bf a}-{\bf h}) Spin structure factor taking into account the atomic positions of {\nab}, calculated in the classical and quantum limits via cMC and PMFRG, respectively. The calculations are performed at two different temperatures, $T=0.5\,J_4$ and $T=0.1\,J_4$, for two different planes: $(q_x,q_y,0)$ and $(q_x,0,q_z)$. All calculations are normalized to the maximum in each panel. ({\bf i}, {\bf j}) Integrated spin structure factor for the three-dimensional models taking into account the positions of the spins and the dimensions of the rectangular unit cell, as well as the magnetic form factor for the Cu$^{2+}$ ions. Blue and green lines show the cMC results at two different temperatures above the critical temperature, while the red curve corresponds to the PMFRG results for the $S=1/2$ case. {\bf i} $S(\mathbf{q})$ integrated only in the $q_xq_y$ plane using the atomic form factor. {\bf j} $S(\mathbf{q})$ integrated in the complete Fourier space using the atomic form factor.}
\label{fig:cmcvspmfrg}
\end{figure*}

We calculate the complete spin structure factor with the real positions of the atoms to allow comparison with future (inelastic) neutron scattering experiments. To take into account all interactions in our cMC calculations as well as the atomic positions in {\nab}, we have to enlarge the unit cell from 7 sites in the 2D model to 56 spins in the 3D case, which includes two layers of $2\times 2$ two-dimensional unit cells. This is needed because consecutive layers in {\nab} have different chirality, meaning that $J_1$ and $J_3$ couplings and the twisting angle of the pyramids are exchanged. At high temperatures, the inter-layer coupling $J_8$ is expected to have a negligible influence on the system and the physics should be the same as for the two-dimensional model. However, the averaging between the two types of layers leads to a more symmetric spin structure factor, as seen in Fig.~\ref{fig:cmcvspmfrg} for $T=0.5\,J_4$ and $0.1\,J_4$ (see also Supplementary Note 7). If the spin structure factor is calculated taking only the spins in one type of layer into account, the spin structure factors for the two-dimensional model are recovered for the two temperatures shown. The spin structure factors on the two types of layers are connected by reflection $q_x \to -q_x$ or $q_y \to q_y$, such that the spin structure factors of the three-dimensional model can be obtained roughly as $S^\text{3D}(q_x,q_y,q_z=0) = S^\text{2D}(q_x,q_y)+S^\text{2D}(-q_x,q_y)$, where $S^\text{2D}(q_x,q_y)$ are equivalent to the results in Fig.~\ref{fig:ssf_cmc} but taking into account the lattice distortions in {\nab} (see Supplementary Figure 11 for more details).

In Fig.~\ref{fig:cmcvspmfrg}, we also compare the results for cMC ($S=\infty$) and PMFRG ($S=1/2$) at two different temperatures above the classical phase transition. The results are presented in a larger region of reciprocal space because the positions of the spins now lead to a spin structure factor which is not periodic in an extended Brillouin zone. In real materials, the form factor of the magnetic atoms causes the spin structure factor to fade for large values of $|\mathbf{q}|$. In Fig.~\ref{fig:cmcvspmfrg}\,{\bf a} to {\bf d}, we show the cMC calculations. On the $xy$ plane, we observe that the spiral surface from the 2D model at $T=0.5\,J_4$ deforms into a horseshoe pattern when considering the spin positions. As the temperature is lowered to $T=0.1\,J_4$, these horseshoes become sharper and evidence the crossover to the needle state which now is averaged over the two chiralities and the weight is shifted by the distortions in the positions of the spins. These horseshoes resemble the square rings observed in the breathing pyrochlore lattice, which originate from the possibility of flipping planes~\cite{Benton15}. We also observe that at low temperatures, a characteristic four-leaf clover is formed in the centre of the reciprocal space. All these features indicate a highly correlated state at finite temperatures, which cannot be understood as the precursor of an ordered state. On the right part of Fig.~\ref{fig:cmcvspmfrg}\,{\bf b}, we show the Bragg peaks structure below the transition temperature (see Fig.~\ref{fig:mag}\,{\bf d} for reference states). This evidences that the spin structure factor features we observe at finite temperatures correspond to the liquid-like features in the two-dimensional model discussed above and are not precursors of the Bragg peaks corresponding to the three-dimensional model. 

In Fig.~\ref{fig:cmcvspmfrg}\,{\bf e} to {\bf h} we show the PMFRG calculations for the quantum $S=1/2$ case for the same two temperatures. In this case, we can see that the horseshoe features survive down to low temperatures ($T=0.1\,J_4$), and the spin structure factor resembles that of the classical case at higher temperatures. We observe a qualitatively similar pattern down to $T=0.01\,J_4$. We do not find a phase transition within PMFRG. The very low critical temperature observed experimentally, of about $0.02\,J_4$, is well below the limit for which finite-temperature phase transitions can be detected reliably with PMFRG. Already at $T=0.1\,J_4$, some dark-blue parts can be seen in Fig.~\ref{fig:cmcvspmfrg}\,{\bf f} coming from negative values in the spin structure factor. These small regions of unphysical values usually appear in low temperatures in PMFRG and should be considered as an artefact. However, the important thing is that the spin structure factor does not change qualitatively down to the smallest temperatures.

Finally, we move to the angle-integrated spin structure factor $S(|\mathbf{q}|=q)$, which can be measured via neutron scattering and is typically used as a first approach to evaluate the ordering tendencies at low temperatures. We show the spin structure factor, integrated over the $q_xq_y$-plane in Fig.~\ref{fig:cmcvspmfrg}\,{\bf i} and integrated over the whole Fourier space in Fig.~\ref{fig:cmcvspmfrg}\,{\bf j}. The first is useful for tracking the origin of the peaks, which originate from the integration circles reaching the different sets of horseshoes from Fig.~\ref{fig:cmcvspmfrg}\,{\bf a} and {\bf b}. The latter is what can be measured experimentally, and shows that these peaks are washed out when taking into account the whole spin structure factor. However, some very characteristic hills and valleys are present both in the classical and quantum calculations at finite temperatures above the phase transition.

\section*{Discussion}

We derived an exchange Hamiltonian for the Cu$^{2+}$ based $S=1/2$ compound {\nab} employing the energy mapping technique within DFT. The Hamiltonian we obtained revealed weakly coupled layers of square-kagome lattices with an extra site at the centre of each square (and only connected to the squares) forming pyramids. Furthermore, the pyramids were shown to have a small twisting angle in the structure that leads, however, to an appreciable difference between the two triangular couplings. While one of them is large and antiferromagnetic, the other one vanishes within error bars. The twisting angle was found to be opposite in consecutive layers, leading to an opposite chirality of interactions (an exchange between triangular bonds from one layer to the neighbouring ones). 

To validate the Hamiltonian we studied the model in a magnetic field employing classical Monte Carlo simulations and found that the total ferrimagnetic moment for the 2D model lies close to the value observed experimentally. Moreover, the 2D model can explain two out of the three slopes experimentally observed in {\nab} and permits us to track the origin of the intermediate-field slope to the sites connecting the pyramids. We also investigated the magnetization process of the 3D model in the classical limit and found an even better agreement with the experimental results from Ref.~\cite{Markina2024_main}, accounting for the extra observed phase transition observed at small magnetic fields. This is a direct consequence of the inclusion of the ferromagnetic inter-layer coupling $J_8$ which, despite being only $\sim 5\%$ of the largest coupling, changes the nature of the ground state completely. Altogether, this demonstrates the essential level of accuracy needed from the DFT Hamiltonian to capture all the important features in the magnetization curve.

Given that the interlayer coupling is small, its effect can be countered in two ways: either by increasing the temperature or by introducing a magnetic field. Both lead to a two-dimensional behaviour. It is therefore interesting, from both theoretical and experimental perspectives, to study the Hamiltonian on single layers. In the classical case, we found several interesting features. To begin with, at moderate temperatures, we observe spiral rings in the total spin structure factor. A closer look at the spin structure factors for different sublattices shows that in this regime, the spins at the centre of the squares are completely paramagnetic, indicating a partially disordered phase. At lower temperatures, these spiral rings break down into needles that indicate very specific directional degrees of freedom. We derived a simplified model Hamiltonian that explains the origin of these needles: they originate from the freedom to exchange spins along specific lines of the system. This leads to a unique type of ferrimagnetic spin liquid at zero temperature. This type of unconventional ferrimagnetism, which emerges solely from antiferromagnetic interactions among spins of equal size, usually appears in systems where individual spins are coupled to a lattice hosting a coplanar order in such a way that they see a zero mean field. Prominent examples are the stuffed honeycomb and square lattices~\cite{Nakano17, Shimada18, Gonzalez19, Blesio23}. In the case of {\nab}, the role is played by the decorating sites over the square-kagome lattice, which are only coupled to the spins in the base of the pyramid by $J_5$ and do not interact directly with one another (i.e., they become orphan spins if $J_5 = 0$).

In the quantum case, we studied the spin structure factor at finite temperatures and found that the high-temperature classical state survives to lower temperatures, while the needle features are never observed. It would be worthwhile to study the same model directly at $T=0$ to check the presence of needle-like features in the ground state of the quantum Hamiltonian. Our analysis also makes it plausible to access interesting ferrimagnetic spin liquid states at finite fields and finite temperatures in nabokoite {\nab}. 

\section*{Methods}
\subsection*{Density Functional Theory-based energy mapping}

We perform DFT calculations using the full potential local orbital (FPLO) basis~\cite{Koepernik1999} and a generalized gradient approximation to the exchange correlation functional~\cite{Perdew1996}. We correct for the strong correlations on the Cu$^{2+}$ ion using a DFT+U correction~\cite{Liechtenstein1995}. The DFT-based energy mapping approach requires calculating precise DFT energies for a large number of distinct spin configurations and fitting them to the classical energies of the Heisenberg Hamiltonian, Eq.~\eqref{eq:H}. In the case of nabokoite, we reduce the symmetry to $P2_1$ in order to make 14 of the 28 Cu$^{2+}$ ions in the unit cell symmetry inequivalent, and we calculate 31 energies which allow us to resolve 13 exchange interactions, among them the six nearest neighbour paths. Further details are in the Supplementary Note 2.

\subsection*{Classical Monte Carlo}

We perform classical Monte Carlo calculations to study the properties of the model Hamiltonian in the classical limit. In this limit, we consider the spins as unit vectors, $|\mathbf{S}|=1$. We simulate systems of $L\times L$ unit cells, consisting of $N=7L^2$ spins for the 2D model, with up to $L=80$ (44800 spins); and of $N=56L^3$ spins for the 3D model, with up to $L=8$ (28672 spins). We apply a logarithmic cooling protocol from $T=2\,J_4$ to $T=0.01\,J_4$ with 120 temperature steps consisting of $10^5$ Monte Carlo steps each. Half of the steps at each temperature are used for thermalization and each consists of $N$ spin-update trials intercalated with $2N$ overrelaxation steps. For the spin updates, a Gaussian step is implemented such that the acceptance ratio is kept at $50\%$~\cite{Alzate19}. The second half of the Monte Carlo steps is used to measure the energy $e$ and the specific heat $c_v$. Other quantities such as the spin structure factor are calculated afterwards from previously stored thermalized states for each temperature. Further details in the Supplementary Notes 3 and 4.

\subsection*{Pseudo-Majorana Functional Renormalization Group}

We use the temperature-flow package for PMFRG which can be found in the GitHub repository~\cite{NilsGit}. The method relies on rewriting the spin $S=1/2$ operators in terms of Majorana fermions. This representation has the advantage of not generating any unphysical states. Only an artificial degeneracy of states is introduced. However, this degeneracy only contributes with a known factor to the free energy and it does not affect the expectation values of the spin-spin correlations~\cite{Shnirman03, Schad16, Muller24}. 

In PMFRG all lattice symmetries (either translational, rotational, or combinations of both) are implemented. Then, all the correlations are calculated between the 3 symmetry inequivalent sites and all the other spins within a certain distance $R$. From the correlations, the spin structure factor can be calculated at any temperature in the flow from $T=\infty$ down to $0.01\,J_4$. Within PMFRG, finite-temperature phase transitions can be detected via finite-size scaling of the correlation length, which can be obtained from the peaks in the spin structure factor~\cite{Niggemann22_main}. In this case, we calculate the peak in the spin structure factor as a function of the temperature, $S^\text{max}(T)$, for different system sizes $R$. In the 2D case, we observe that $S^\text{max}(T)$ does not change above $R=10$, where 1 is the nearest-neighbour distance, down to $T=0.01$. This indicates that there is no finite-temperature phase transition and that calculating the correlation length would not lead to any crossings for different system sizes (no finite-size scaling is possible or needed). For the 3D model, we observe that $S^\text{max}(T)$ does not change above $R=10$~{\AA}, where the nearest-neighbour distance is 3.102~{\AA}. This again means that there is no finite-temperature phase transition. More information about the method and the flows can be found in Supplementary Note 5.

\section*{Data Availability}

All the data that support the findings of this study are available in a Zenodo repository~\cite{Zenodo}.

\section*{Code Availability}

The DFT code used in this study is available from \texttt{https://www.fplo.de/}. The code for the cMC calculations of this study is available from the corresponding author upon request. The PMFRG code can be found in a GitHub repository~\cite{NilsGit}.

\section*{Acknowledgments}

We thank M. M. Markina for sharing the data from Ref.~\cite{Markina2024_main} which is included in Fig.~\ref{fig:mag}\,{\bf c}. We thank A. Vassiliev, S. Streltsov, S. Trebst, N. Niggemann and K. Penc for helpful discussions and collaboration on related projects. We acknowledge the use of the JUWELS cluster at the Forschungszentrum J{\"u}lich and the HPC Service of ZEDAT, Freie Universit{\"a}t Berlin. J. R. acknowledges support from the Deutsche Forschungsgemeinschaft (DFG, German Research Foundation), within Project-ID 277101999 CRC183 (Project A04). H.~O.~J. acknowledges support through JSPS KAKENHI Grant No. 24H01668. The work of Y.~I. and J.~R. was performed, in part, at the Aspen Center for Physics, which is supported by National Science Foundation Grant No.~PHY-2210452. The participation of Y.I. at the Aspen Center for Physics was supported by the Simons Foundation (1161654, Troyer). This research was supported in part by grant NSF PHY-2309135 to the Kavli Institute for Theoretical Physics (KITP) and the International Centre for Theoretical Sciences (ICTS), Bengaluru through participating in the program - Kagome off-scale (code: ICTS/KAGOFF2024/08). Y.~I.\ acknowledges support from the ICTP through the Associates Programme, from the Simons Foundation through Grant No.~284558FY19, and IIT Madras through the Institute of Eminence (IoE) program for establishing QuCenDiEM (Project No.~SP22231244CPETWOQCDHOC). Y.~I.\ also acknowledges the use of the computing resources at HPCE, IIT Madras. J.R. and H.O.J. thank IIT Madras for a Visiting Faculty Fellow position under the IoE program during which this work was initiated and completed.

\section*{Author contributions}

The project was conceived by HOJ and YI. HOJ performed the DFT and energy-mapping calculations. MGG performed the cMC and PMFRG calculations. MGG, YI, JR, and HOJ analysed and discussed the results, and wrote and reviewed the manuscript.

\section*{Competing interests}

The authors declare no competing interests.


%

\clearpage
 
\beginsupplement

\onecolumngrid

\begin{center}
{\bf \large Supplementary Information for ``Field-induced spin liquid in the decorated square-kagome antiferromagnet nabokoite {\nab}"}\\[1.5em]

Mat\'ias G. Gonzalez,$^{1,2}$ Yasir Iqbal,$^3$ Johannes Reuther,$^{1,2,3}$ and Harald O. Jeschke,$^{4,3}$\\[0.5em]

\textit{\small
$^1$Helmholtz-Zentrum Berlin f\"ur Materialien und Energie, Hahn-Meitner-Platz 1, 14109 Berlin, Germany\\
$^2$Dahlem Center for Complex Quantum Systems and Fachbereich Physik, Freie Universität Berlin, 14195 Berlin, Germany\looseness=-1\\
$^3$Department of Physics and Quantum Centre of Excellence for Diamond and
Emergent Materials (QuCenDiEM),\\
Indian Institute of Technology Madras, Chennai 600036, India\\
$^4$Research Institute for Interdisciplinary Science, Okayama University, Okayama 700-8530, Japan
}
\end{center}

\twocolumngrid

\begin{figure}[htb]
\centering
\includegraphics[width=\columnwidth]{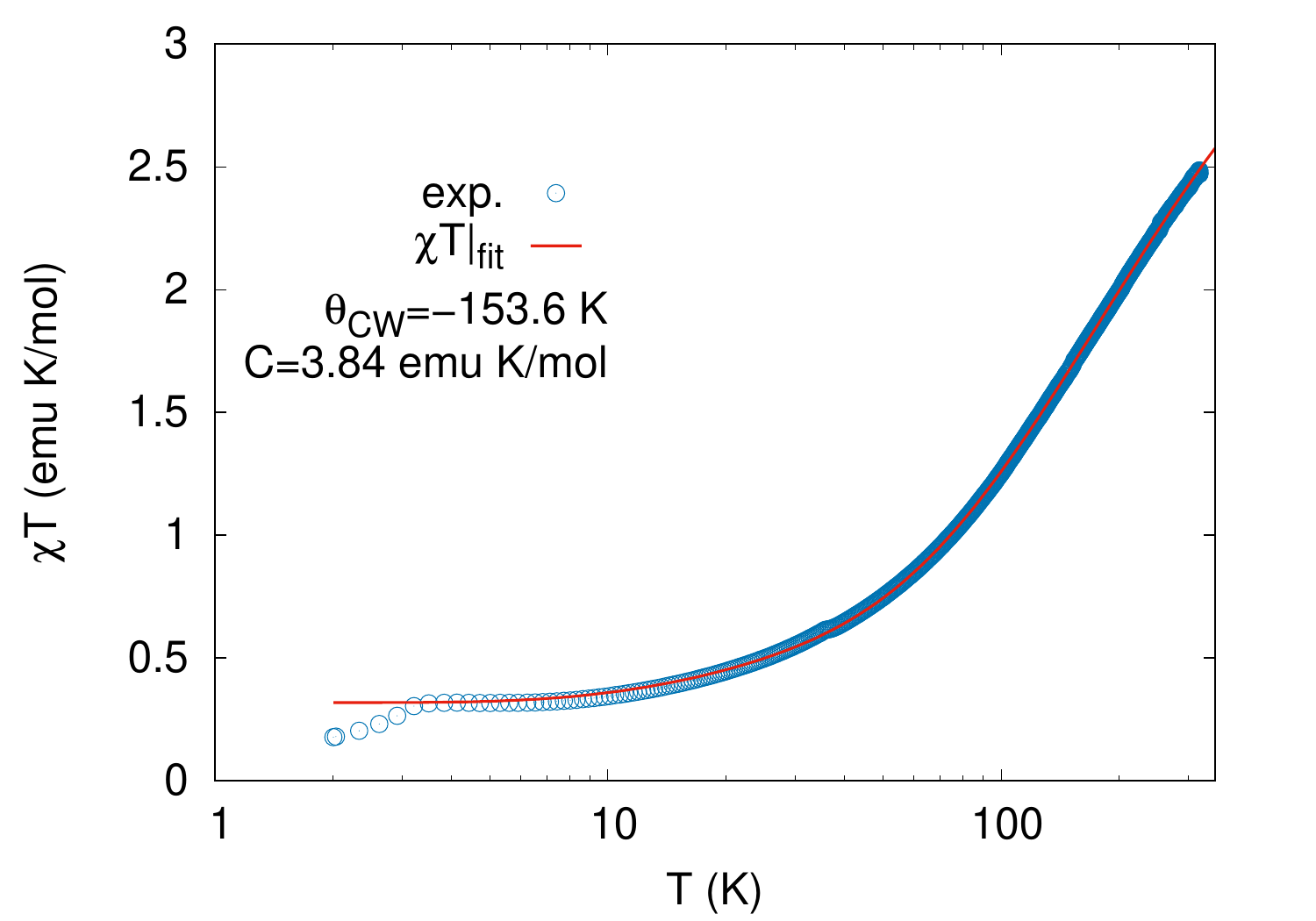}
\caption{Magnetic susceptibility $\chi(T)$ of nabokoite from Ref.~\cite{Markina2024}, multiplied by temperature (symbols) and fit by the {\it ansatz} of Eq.~\eqref{eq:ansatz}.}
\label{fig:chiT}
\end{figure}

\section{Re-evaluating the experimental susceptibility}\label{sec:cw}

Nabokoite {\nab} is a highly frustrated antiferromagnet, with ordering temperature $T_{\rm N}=3.2$\,K strongly suppressed below the energy scale of magnetic interactions. We thus use a fitting {\it ansatz} developed recently by Pohle and Jaubert~\cite{Pohle2023} for spin liquids:
\begin{equation}
   \begin{split}
    &\chi T\big|^{\rm fit} =\frac{1+b_1\exp[c_1/T]}{a+b_2\exp[c_2/T]}\\
    &C=\frac{1+b_1}{a+b_2},\quad \theta_{\rm CW}=\frac{b_1c_1}{1+b_1}-\frac{b_2c_2}{a+b_2}
\end{split}\label{eq:ansatz}
\end{equation}
The fit is shown in Supplementary Figure~\ref{fig:chiT}. It is excellent right down to the ordering temperature $T_{\rm N}$ and allows us to determine the Curie-Weiss temperature of nabokoite to $\theta_{\rm CW}=-153.6$\,K. Note that a linear fit of $\chi^{-1}$ is not feasible for this materials~\cite{Markina2024}.

\section{Additional DFT details}\label{sec:dft}

\begin{figure}[!t]
\centering
\includegraphics*[width=\columnwidth]{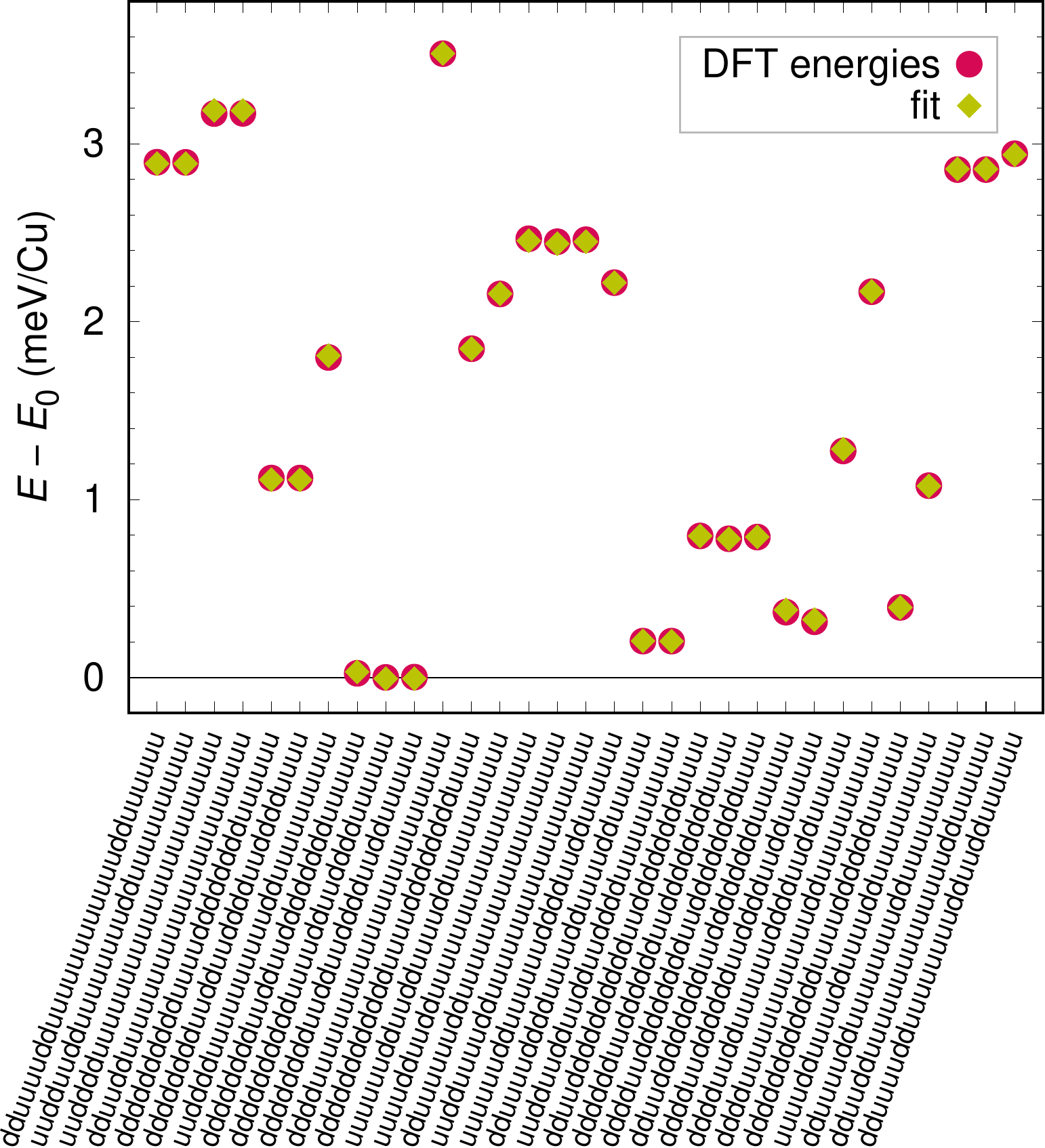}
\caption{DFT total energies per Cu for 31 distinct spin configurations (circles). The fit to the Heisenberg Hamiltonian with 13 exchange interactions is excellent.}
\label{fig:energies}
\end{figure}

\begin{table*}[htb]
\centering
\caption{Exchange interactions of nabokoite {\nab} obtained by DFT energy mapping as described in the Methods section. The line in bold face corresponds to the set of couplings that match the experimental Curie-Weiss temperature. The distances $d$ given in the last line are the Cu-Cu distances that identify the exchange paths.}\label{tab:couplings}
  \begin{tabular}{c|c|c|c|c|c|c|c}
  $U$\,eV & $J_1 $\,(K) & $J_2 $\,(K) & $J_3 $\,(K) & $J_4 $\,(K) & $J_5 $\,(K) & $J_6 $\,(K) & $J_8 $\,(K) \\\hline
 5 & 25.3(6.7) & 212.3(4.3) & 248.3(4.8) & 255.5(4.9) & 241.0(2.3) & 43.2(2.7) & -14.4(1.7) \\
5.5 & 19.4(5.3) & 189.3(3.5) & 229.7(3.9) & 236.3(4.0) & 222.8(1.9) & 40.5(2.2) & -13.2(1.3) \\
6 & 14.3(4.3) & 168.8(2.8) & 212.2(3.1) & 218.4(3.2) & 204.3(1.5) & 37.9(1.7) & -12.2(1.1) \\
7 & 5.4(2.8) & 134.1(1.8) & 179.9(2.1) & 186.1(2.0) & 170.1(1.0) & 33.1(1.2) & -10.6(0.7) \\
7.5 & 2.7(2.2) & 119.8(1.5) & 166.5(1.6) & 171.8(1.7) & 154.0(0.8) & 30.7(0.9) & -9.6(0.6) \\
{\bf 7.55} & {\bf 2.4(1.8)} & {\bf 118.5(1.5)} & {\bf 165.2(1.6)} & {\bf 170.5(1.7)} & {\bf 152(0.8)} & {\bf 30.5(0.9)} & {\bf -9.5(0.6)} \\
8 & -0.5(1.8) & 106.7(1.2) & 152.8(1.3) & 158.4(1.3) & 140.0(0.7) & 28.6(0.8) & -9.0(0.5)\\\hline
$d$\,({\AA})& 3.102 & 3.294 & 3.453 & 4.659 & 4.672 & 4.917 & 5.366 
\\\hline\\\hline
 $U$\,eV & $J_{9}$\,(K) & $J_{11}$\,(K) & $J_{12}$\,(K) & $J_{16}$\,(K) & $J_{18}$\,(K) & $J_{20}$\,(K) & $\theta_{\rm CW}$\,(K)\\\hline
5 & 2.3(4.2) & -17.6(4.7) & -5.9(5.3) & -2.4(4.1) & 0.1(2.4) & 11.2(3.4) & -249 \\
5.5 & 1.9(3.5) & -15.2(3.7) & -4.7(4.3) & -2.4(3.3) & 0.1(1.9) & 8.9(2.8) & -228 \\
6 & 1.6(2.7) & -13.2(3.0) & -3.8(3.4) & -2.2(2.6) & 0.1(1.5) & 7.2(2.2) & -207 \\
7 & 0.8(1.8) & -9.8(2.0) & -3.2(2.2) & -1.9(1.7) & 0.1(1.1) & 4.9(1.4) & -171 \\
7.5 & 0.9(1.4) & -9.0(1.6) & -2.3(1.8) & -1.8(1.4) & 0.1(0.8) & 3.9(1.2) & -155 \\
{\bf 7.55} & {\bf 0.9(1.4)} & {\bf -8.9(1.6)} & {\bf -2.2(1.8)} & {\bf -1.8(1.4)} & {\bf 0.1(0.8)} & {\bf 3.9(1.2)} & {\bf -153.6} \\
8 & 0.6(1.2) & -7.8(1.3) & -2.3(1.5) & -1.6(1.1) & 0.0(0.7) & 3.4(0.9) & -140 \\\hline
$d$\,({\AA})&  5.678 & 5.693 & 5.892 & 6.204 & 6.851 & 6.953 & \\\hline
\end{tabular}
\end{table*}

In Supplementary Figure~\ref{fig:energies}, we show a comparison between the DFT energies and the fit to the Heisenberg Hamiltonian with 13 exchange interactions. The agreement is very good. Small deviations are reflected in the statistical errors given in Supplementary Table~\ref{tab:couplings} which contains the full energy mapping results. Exchange paths are identified by the Cu-Cu distances given in the last line. The first six exchange interactions are shown graphically in Fig.~1\,{\bf a} of the main text. The bold line in the table is the interpolated set of couplings which matches the Curie-Weiss temperature of $\theta_{\rm CW}=-153.6$\,K. We calculate $\theta_{\rm CW}$ according to
\begin{equation}\begin{split}
    \theta_{\rm CW}& =-\frac{2}{3}S(S+1)\frac{2}{7}\big(2J_1 + 2J_2 + 2J_3 + J_4 + 2J_5 + 2J_6 \\ &+ 2J_8+ 2J_9+ 2J_9 +  2J_{11} + 2J_{12} + J_{16} + J_{18} + 2J_{20} \big)
\end{split}\end{equation}

\section{Classical Monte Carlo results on the 2D models}

The classical Monte Carlo (cMC) calculations are carried out as explained in the Methods section of the main article. For the 2D model, we use system sizes of $N=7 L^2$, where 7 is the number of sites in the unit cell and $L$ is the number of unit cells in the two Cartesian directions. In each independent run, the energy and specific heat are calculated at every temperature by averaging through the second half of the Monte Carlo steps performed at each temperature. While the energy is calculated by a simple average, $e=\langle e \rangle$, the specific heat is calculated as
\begin{equation}
    c_v(T) = N \frac{\langle e^2 \rangle - \langle e \rangle^2 }{T^2}
\end{equation}
The results are then averaged over 10 independent runs. The $c_v(T)$ calculations (after averaging) are shown in Supplementary Figure~\ref{fig:2Dmodel}\,{\bf a}, by the continuous lines for different system sizes from $L=10$ ($N=700$ spins) to $L=80$ ($N=44800$ spins). 

Another alternative is to calculate $c_v(T)$ as the derivative of $e(T)$ after averaging over independent runs. The results of this method are shown by the dashed lines in Supplementary Figure~\ref{fig:2Dmodel}\,{\bf a}, where an arbitrary 0.3 shift has been chosen to distinguish from the previous results. Although the two methods for obtaining $c_v(T)$ should be equivalent, one often encounters discrepancies. When this happens, it is usually an indication that the system is exhibiting problems in thermalizing. In this case, both curves agree well down to the lowest temperatures, and only differ slightly around the peak at $T=0.004\,J_4$. This peak is very faint and does not scale with system size, indicating that it is not a phase transition. Furthermore, it seems to vanish with increasing system size. And since the continuous SU(2) symmetry cannot be broken at finite temperatures in 2D systems, a phase transition (if there is such a transition) has to come from the breaking of an emergent discrete symmetry, none of which is broken in the ground state.

\begin{figure}[!t]
\centering
\includegraphics*[width=\columnwidth]{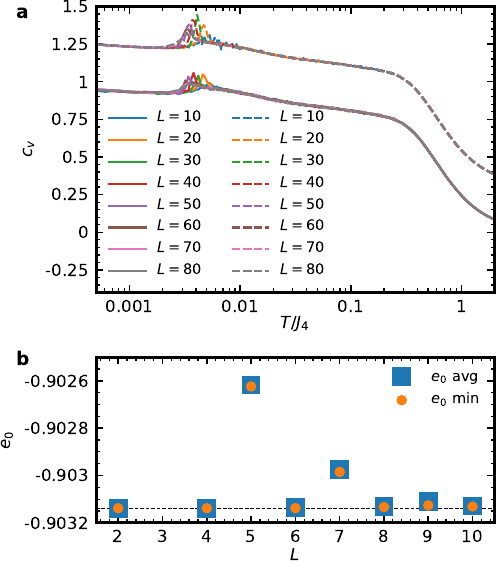}
\caption{Classical Monte Carlo calculations for the 2D model for several sizes $L$. {\bf a} Specific heat is calculated in two different ways (see text), where the dashed data is shifted by 0.3 to appreciate the difference. {\bf b} Ground-state energy average over 10 independent runs (blue squares) and the lowest value among the runs (orange circle).}
\label{fig:2Dmodel}
\end{figure}

The ground-state energy of the system can be obtained as the continuation of a slow cooldown protocol to $T=0$. In Supplementary Figure~\ref{fig:2Dmodel}\,{\bf b} we show these results for smaller lattices ranging from $L=2$ to 10. The average ground-state energy is shown by the blue squares, while the orange dots show the minimum among the 10 independent runs for each lattice size. It becomes evident that all lattices with even $L$ reach the same ground-state energy, even for lattices as small as $L=2$ ($N=28$ spins). On the other hand, the odd $L$ lattices exhibit higher energies ($L=3$ is out of scale), indicating that the periodic boundary conditions are introducing frustration into the system, while the even $L$ lattices are unfrustrated.

As explained in the main text, this behaviour can be tracked to the spin pattern of the ground state, which requires 2 unit cells in each direction to repeat itself along independent zigzag lines of $J_3$ and $J_5$. We show in Supplementary Figure~\ref{fig:2Dmodelconf} two examples of ground-state configurations for the $L=2$ lattice. The arrows indicate the $(S_x, S_y)$ components of the spins, while the colour determines $S_z$. For clarity, we rotate all the spins such that the bottom left apex spin is $\mathbf{S} = (0,0,1)$; therefore the green arrow with zero length. The direction of this arrow, as well as the other apex arrows, should be disregarded since it is just selected by a small numerical error in the $(S_x,S_y)$ components. The important thing is that they are all dark green and have almost zero length.

These states have already been described in the main text, but it is nonetheless important to revisit it's features. All the base spins are slightly canted out of the $(S_x,S_y)$-plane and have a small negative $S_z$ component, as can be seen by their yellowish-red colour. The link spins are also canted, but with positive values of $S_z$. Overall, base and link spins have only four different directions in the $(S_x,S_y)$-plane. The base spins connected by diagonal couplings $J_4$ (orange) are antiferromagnetically ordered in $(S_x, S_y)$, while they form $\pi/2$ angles with neighbouring spins connected by $J_2$ (purple). The base spins are then defined by pointing opposite [in $(S_x, S_y, S_z)$] to their neighbours connected by $J_3$ (red coupling). 

Altogether, this is the four-coloured solution presented in the main text. The spins on zigzag lines (see main text) composed of base and link spins connected by $J_4$ and $J_3$ (orange and red bonds) can be transformed according to $(S_x, S_y, S_z) \to (-S_x, -S_y, S_z)$ while maintaining the same energy, staying in the ground-state manifold. This gives rise to the subextensive degeneracy since there are $2L$ lines in which these moves can be applied independently. When applying cMC to larger lattices, as the temperature is lowered, the system gets trapped into a particular state of the manifold which, however, is spatially diverse in the sense that it does not contain a repeated $L=2$ cell. Therefore, large enough lattices represent the variety within the manifold well and give rise to the needle-like features in the spin structure factor (see main text).

\begin{figure}[!t]
\centering
\includegraphics*[width=\columnwidth]{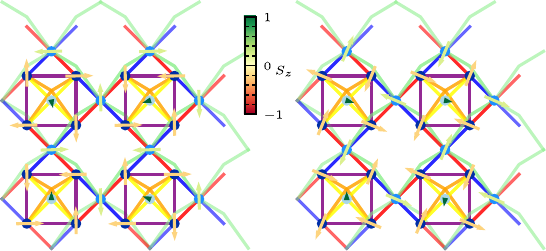}
\caption{Two classical Monte Carlo configurations for the 2D model at $T=0$ for $L=2$. The size and direction of the arrows indicate the $(S_x,S_y)$ component, while the colour of the arrows indicates the $S_z$ component. All spins are rotated so that the appex spin in the bottom left is $(0,0,1)$.}
\label{fig:2Dmodelconf}
\end{figure}

As shown in the main article, the needle-like features in the spin structure factor can be reproduced in an effective model in which 2/7 of the spins are missing and $J_1$ and $J_6$ couplings are neglected while $J_3$ changed sign (from antiferromagnetic to ferromagnetic). The number of sites for these systems is $N=5 L^2$, because the unit cell now contains only five sites (one pyramid). In Supplementary Figure~\ref{fig:2Deff} we show the results for this effective 2D model. In this case, in contrast to the full 2D model, there is no peak at finite temperatures. On the energy side, we see again that even $L$ always gets the same and lowest ground-state energies. On the other hand, odd $L$ reaches higher energies for $L=3$ and $L=5$ but is indistinguishable from even $L$ for $L\geq 7$. This implies that the frustration induced by the periodic boundary conditions is quickly released in larger system sizes.

\begin{figure}[!t]
\centering
\includegraphics*[width=\columnwidth]{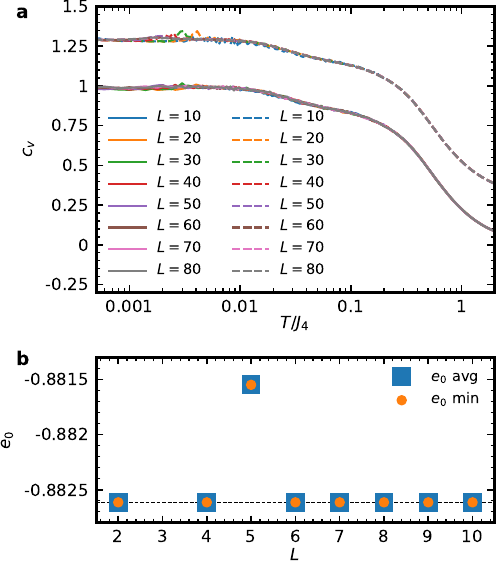}
\caption{Classical Monte Carlo calculations for the 2D effective model for several sizes $L$. {\bf a} Specific heat is calculated in two different ways (see text), where the dashed data is shifted by 0.3 to appreciate the difference. {\bf b} Ground-state energy average over 10 independent runs (blue squares) and the lowest value among the runs (orange circle).}
\label{fig:2Deff}
\end{figure}

While both models show the same key features in the spin structure factor, their specific heats differ in the presence or absence of a peak. The source of the difference becomes evident when looking at the ground-state configuration, shown in Supplementary Figure~\ref{fig:2Deffconf}. The basic pattern is similar to the one in the full 2D model, with all apex spins pointing in the $S_z$ direction and the base spins canted out of the $(S_x, S_y)$-plane by a small negative value of $S_z$. The spins diagonal to each other, connected by the orange bond $J_4$, have also opposite values of $(S_x, S_y)$. The only difference in the effective model is that now all zigzag lines are independent, meaning that the spins connected by the purple $J_2$ bonds can form any angle (and not $\pi / 2$ as before); see in Supplementary Figure~\ref{fig:2Deffconf} that they form different angles in different squares. This degeneracy is lifted in the full 2D model by the bonds $J_1$, $J_6$, and the inclusion of the link sites. This explains the peak observed in the specific heat, corresponding to the temperature at which the angle is locked to $\pi / 2$ when the small couplings become relevant.

\begin{figure}[!t]
\centering
\includegraphics*[width=\columnwidth]{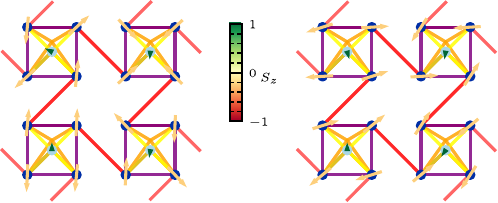}
\caption{Two classical Monte Carlo configurations for the effective model at $T=0$ for $L=2$. The size and direction of the arrows indicate the $(S_x, S_y)$ component, while the colour of the arrows indicates the $S_z$ component. All spins are rotated so that the apex spin in the bottom left is $(0,0,1)$.}
\label{fig:2Deffconf}
\end{figure}

In conclusion, the effective model has an extra freedom, in which all the spins along a certain zigzag line can be rotated by any angle around the $S_z$ axis and not only by an angle of $\pi$ as in the full 2D model. This means that the four colouring solutions are still contained within the manifold in the effective model, but the space is larger. However, both systems exhibit the characteristic needle-like features in the spin structure factor because it only depends on the possibility of the system to fluctuate freely along zigzag lines.

\section{Classical Monte Carlo results on the 3D model}

To consider the 3D model that describes nabokoite, we take into account the lattice distortions that affect the structural periodicity. Because on each layer there are pyramids pointing up and down, we take as the layer unit cell an $L=2$ piece of the 2D model. Furthermore, consecutive layers have exchanged $J_1$ and $J_3$ couplings due to the pyramids being slightly rotated clockwise or counterclockwise. Therefore, we need to consider as the unit cell of our 3D system two layers of $L=2$ in the language of the 2D model, equivalent to $N_u = 7\times 2^3 = 56$ spins. With this in mind, we make our calculations in systems of $N=56 L^3$ spins with up to $L=8$ (28672 spins). The rest of the calculations are performed in the same way as for the 2D models.

In Supplementary Figure~\ref{fig:3Dmodel}\,{\bf a} we show the results for the specific heat calculated in two different ways. Both methods agree in the sense that they exhibit a peak that scales with increasing system size, showing a clear phase transition. Furthermore, we also plot the specific heat for the $L=40$ 2D model in a dot-dashed black line. It becomes clear that the calculations for the 3D and 2D models start to be different below $T=0.2~J_4$ when the interlayer coupling $J_8$ becomes relevant. In Supplementary Figure~\ref{fig:3Dmodel}\,{\bf b} we show the ground-state energies for different sizes. In this case, all lattices with even $L$ exhibit lower ground-state energies and always the same value as the $L=2$ system (448 spins). On the other hand, lattices with odd values of $L$ have larger ground-state energies, indicating that there is frustration due to boundary conditions.

\begin{figure}[!t]
\centering
\includegraphics*[width=\columnwidth]{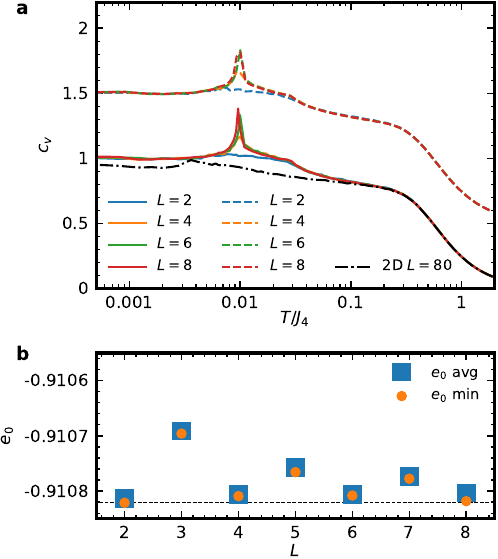}
\caption{Classical Monte Carlo calculations for the 3D model for several sizes $L$. {\bf a} Specific heat is calculated in two different ways (see text), where the dashed data is shifted by 0.5 to appreciate the difference. {\bf b} Ground-state energy average over 10 independent runs (blue squares) and the lowest value among the runs (orange circle).}
\label{fig:3Dmodel}
\end{figure}

In Supplementary Figure~\ref{fig:3Dmodelconf} we show the spin configuration for the $L=2$ lattice using the 3D model. This system contains four different layers, which can be seen on the different panels. As shown in the main article, the ground state of the 3D model is different from that of the 2D model. The interlayer coupling $J_8$, however small, plays a key role at low temperatures driving a phase transition to a more complicated state. The apex spins are not all ferromagnetically ordered through the system. However, there are some slices in which they (the apex spins) order ferromagnetically (see for example the top row on each layer). Some other slices oppose these spins (see the third row from the top). The complicated structure that minimizes the ground-state energy at $T=0$ certainly deserves further investigation, which is out of the scope of this work in which we focus on the properties of the 2D regime.

\begin{figure*}[!t]
\centering
\includegraphics*[width=\textwidth]{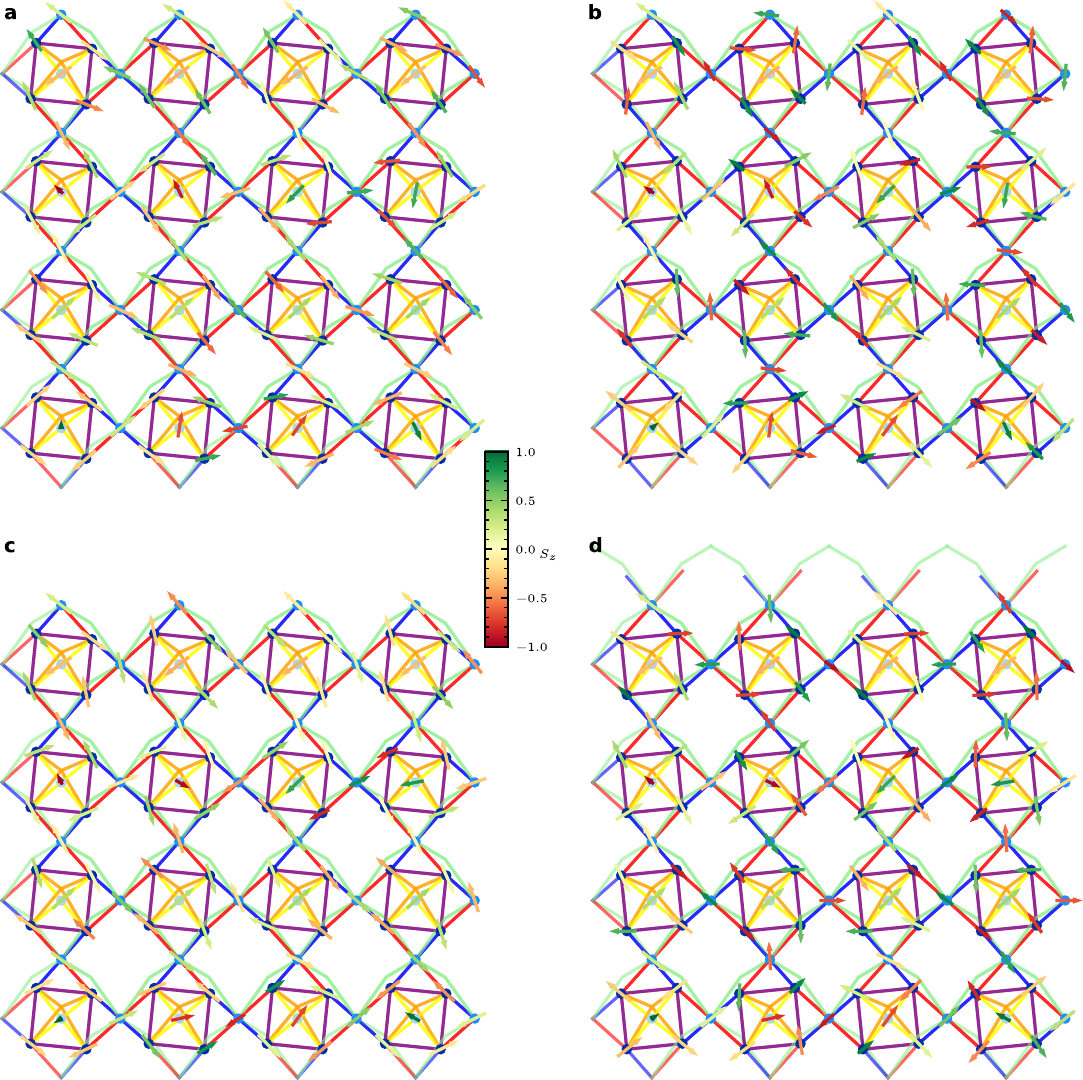}
\caption{Classical Monte Carlo configuration for the 3D model at $T=0$ for $L=2$. The layers 1 to 4 are shown in panels {\bf a} to {\bf d}, respectively. The size and direction of the arrows indicate the $(S_x, S_y)$ component, while the colour of the arrows indicates the $S_z$ component. All spins are rotated so that the apex spin in the bottom left of the first layer is $(0,0,1)$.}
\label{fig:3Dmodelconf}
\end{figure*}

\section{PMFRG results}

The PMFRG method relies on writing the quantum spin operators in the Majorana fermion representation~\cite{Niggemann21, Niggemann22}. This re-writing has the advantage of not introducing any unphysical enlargement of the Hilbert space, but a trivial degeneracy that can be easily considered. In particular, we use the recently developed temperature-flow PMFRG~\cite{Schneider24}, in which the temperature $T$ is used as the cutoff parameter which evolves from $T=\infty$ down to lower temperatures. Essentially, the method consists of solving an infinite set of coupled differential equations, known as the flow equations, for the corresponding fermionic vertex functions, from which the spin-spin correlations can be obtained at a given $T$.

The PMFRG method preserves all symmetries of the original Hamiltonian, and the lattice symmetries can be implemented to reduce considerably the number of flow equations. Within the method, ordering phase transitions at finite temperatures can be detected via finite-size scalings of the correlation length. The latter is obtained directly from the peak in the spin structure factor (or susceptibility). PMFRG has been shown to obtain accurate results for the critical temperature when compared against quasi-exact methods like quantum Monte Carlo~\cite{Niggemann22, Schneider24}.

\begin{figure}[!t]
\centering
\includegraphics*[width=\columnwidth]{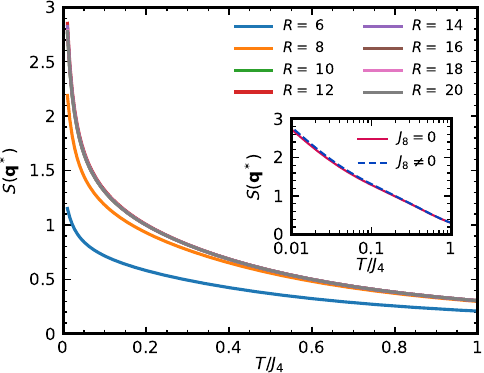}
\caption{PMFRG temperature flows for the 3D model for different values of $R$ (in {\AA}). For each case, $\mathbf{q}^*$ corresponds to the point in reciprocal space for which the highest value of the spin structure factor is observed. The inset shows the difference between the 3D model and taking $J_8=0$ for $R=20$.}
\label{fig:pmfrg}
\end{figure}

To solve the 3D model, we use the available atomic positions found in the literature, for which the nearest-neighbour distances are stated in Table~\ref{tab:couplings}. We take into account couplings from $J_1$ to $J_8$, which correspond to interatomic distances of $3.102$~{\AA} to $5.366$~{\AA}. We solve the flow equations for all symmetry-inequivalent spin-spin correlations contained within a sphere of radius $R$ from the three symmetry-inequivalent Cu sites. In other words, we work on an infinite lattice where we take into account all spin-spin correlations at distances $\leq R$. From these correlations, the spin structure factor can be calculated and subsequently the susceptibility as the values of $S(\mathbf{q}^*;T)$, where $\mathbf{q}^*$ is the wavevector for which the maximum of $S(\mathbf{q})$ can be found.

We show in Supplementary Figure~\ref{fig:pmfrg} the susceptibility in the 3D model as a function of the temperature for several values of $R$ from 6~{\AA} to 20~{\AA}. These imply taking into account between 27 and 823 symmetry-inequivalent correlation pairs. Firstly, no divergency is observed in the temperature range down to $T=0.01~J_4$, indicating the absence of a phase transition. Furthermore, results are converged for all $R\geq 10$, again indicating that the correlation length does not diverge down to $T=0.01~J_4$. This does not rule out the existence of a phase transition at lower temperatures, which are not reliably accessible within PMFRG. We also show in the inset the difference between setting the interlayer coupling to zero, $J_8=0$, for $R=20$~{\AA}. The difference that $J_8$ induces in the susceptibility or spin structure factor is only very small at very small temperatures in the quantum case.

\section{Single classical pyramid}

In this section, we focus on a single pyramid and ask how the energy can be minimized in such a 5-spin system. We consider the base spins connected with $J_2$ along the sides and $J_4$ along the diagonals, while the apex spin is connected by $J_5$ to the base (the same as in all of our models). Even though every spin in the pyramid interacts with the remaining ones, because of the difference in the couplings $J_2 \neq J_4$, the corresponding 5-spin Hamiltonian cannot be rewritten as a complete square. 

First, let us think about the four Cu(1) spins connected in a square with $J_2$ on the sides and $J_4$ on the diagonals, disregarding the apex site and the $J_5$ coupling. This square has two possible solutions depending on the ratio of antiferromagnetic $J_2$ and $J_4$. On one hand, if the diagonal coupling $J_4$ is smaller than $J_2$, the spins minimize the energy by forming a colinear N\'eel state in the square. On the other, if $J_4 > J_2$ as in the present case, the energy is minimized by a coplanar state in which each diagonal has opposing spins, but the angle between spins in different diagonals is completely free.

Let us assume that the coplanar state on the square is on the $xy$ plane, and let us couple the apex spin pointing in the $z$ direction to the square by increasing $J_5$. For $J_5 \to 0$, the state on the square will remain coplanar. When $J_5$ increases, the spins on the square will start canting outside the plane, leading to a total net ferrimagnetic moment. The out-of-plane canting angle is shown in Supplementary Figure~\ref{fig:canting} as a function of $J_5$ and $J_4$. The red star marks the ratio of parameters from the Hamiltonian, and the black contour lines indicate three angles close to it. The dashed black line separates the two states described above, also depicted by the squares and spins in Supplementary Figure~\ref{fig:canting}. As $J_5 \to \infty$, both states become the same: the centre spin pointing up and all the rest pointing down.

\section{From the 2D to the 3D spin structure factor}

Here we illustrate how the spin structure factor observed in the simple 2D model without lattice distortions looks like when considering different layers with opposite chirality in the 3D model as well as lattice distortions. In Supplementary Figure~\ref{fig:factors} we show cMC results for the spin structure factor at two different temperatures, $T=0.5~J_4$ and $T=0.1~J_4$, in two different rows. The first column contains the same results shown in the main article for the 2D model, where the lattice is simplified with respect to the structure of the {\nab} compound. This is done to work with a periodic spin structure factor and with square sublattices in the structure, which in time allows us to interpret more easily the results. 

\begin{figure}[!t]
\centering
\includegraphics*[width=\columnwidth]{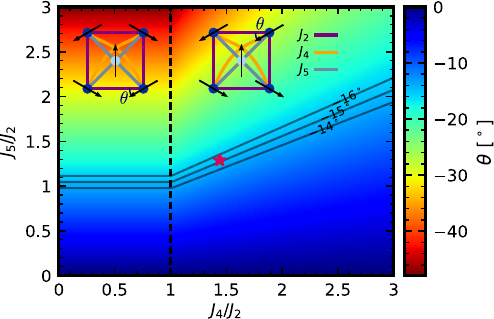}
\caption{Canting angle as a function of $J_4$ and $J_5$ for a single classical pyramid. The lattices at each side of the dashed line depict the states. The red star indicates the values for the {\nab} compound.}
\label{fig:canting}
\end{figure}

\begin{figure*}[b]
\centering
\includegraphics*[width=\textwidth]{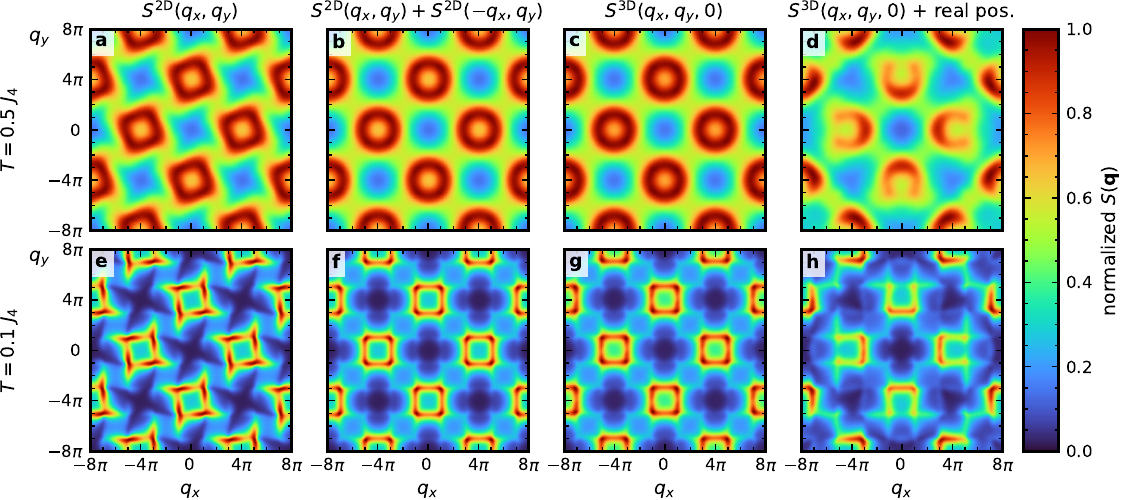}
\caption{Spin structure factor obtained with cMC for two different temperatures, $T=0.5~J_4$ (a-d) and $T=0.1~J_4$ (e-h). The first column (a,e) corresponds to the 2D model, and the second (b-f) also corresponds to the 2D model but is symmetrized to consider the two chiralities of successive layers. The third column (c,g) corresponds to the 3D model without taking into account the real positions of the spins in {\nab}. The last column (d,h) considers the atomic positions.}
\label{fig:factors}
\end{figure*}

In the second column (panels b and f), we show the result of symmetrizing the spin structure factor to take into account different layers with opposing chirality of couplings (exchanging $J_1$ and $J_3$ triangular couplings). Another interpretation is that this is the structure factor that would be observed if the layers are completely decoupled or if $J_8=0$. In the third column (panels c and g) we show the actual calculations on the 3D model, which make evident that $J_8$ is not playing any role at these temperatures and the spin structure factors are the same as in the 2D model. In other words, there is a two-dimensionalization effect led by the temperature. Finally, in the last column, we show the spin structure factor of the 3D model taking into account the atomic positions of the Cu atoms. This shifts the weight in the spin structure factor and makes it non-periodic (as can be seen in the main article). It can be seen that the change in the positions within the unit cell redistributes the weight on the rings, transforming them into the horseshoe features shown in the main article. However, we stress that the atomic positions only change the appearance of the spin structure factor and not the spin-spin correlations themselves, which only depend on the Hamiltonian (that is unchanged).

\clearpage

\end{document}